# Magnetization oscillations and waves driven by pure spin currents

V.E. Demidov[1], S. Urazhdin[2], G. de Loubens[3], O. Klein[4], V. Cros[5], A. Anane[5], and S.O. Demokritov[1*]

[1]*Institute for Applied Physics and Center for Nanotechnology, University of Muenster, Corrensstrasse 2-4, 48149 Muenster, Germany*

[2]*Department of Physics, Emory University, Atlanta, GA 30322, USA*

[3]*Service de Physique de l' Etat Condense, CEA Saclay, 91191 Gif-sur-Yvette, France*

[4]*INAC-SPINTEC, CEA/CNRS and Univ. Grenoble Alpes, 38000 Grenoble, France*

[5]*Unité Mixte de Physique CNRS, Thales, Univ. Paris Sud, Université Paris-Saclay, 91767 Palaiseau, France*

*Corresponding author, e-mail: demokrit@uni-muenster.de

## Abstract

Recent advances in the studies of pure spin currents – flows of angular momentum (spin) not accompanied by the electric currents – have opened new horizons for the emerging technologies based on the electron's spin degree of freedom, such as spintronics and magnonics. The main advantage of pure spin current, as compared to the spin-polarized electric current, is the possibility to exert spin transfer torque on the magnetization in thin magnetic films without electrical current flow through the material. In addition to minimizing Joule heating and electromigration effects, this characteristic enables the implementation of spin torque devices based on the low-loss insulating magnetic materials, and offers an unprecedented geometric flexibility. Here we review the recent experimental achievements in investigations of magnetization oscillations excited by pure spin currents in different magnetic nanosystems based on metallic and insulating magnetic materials. We discuss the spectral properties of spin-current nano-oscillators, and relate them to the spatial characteristics of the excited dynamic magnetic modes determined by the spatially-resolved measurements. We also show that these systems support locking of the oscillations to external microwave signals, as well as their mutual synchronization, and can be used as efficient nanoscale sources of propagating spin waves.

**Keywords:**

Magnetization dynamics, spin waves, spin current, magnonics, magnetic nanostructures

**Contents**

# 1. Introduction

Since the first demonstration [1-5] of the possibility to excite magnetization dynamics by spin-polarized electric currents due to the spin transfer torque (STT) effect [6,7], dynamic spin torque phenomena have become the subject of intense research [8-15]. The ability to control high-frequency magnetization dynamics by dc currents is promising for the generation of microwave signals [8-15] and propagating spin waves [16-23] in magnetic nanocircuits. Electronically-controlled local generation of spin waves is particularly important for the emerging field of nanomagnonics, which utilizes propagating spin waves as the medium for the transmission and processing of signals, logic operations, and pattern recognition on nanoscale [20,24-30].

STT phenomena have been traditionally studied in nanodevices based on the tunneling or giant magnetoresistance spin-valve structures, where STT is induced by the electric current flowing through a multilayer that consists of a "fixed" magnetic spin-polarizer and the active magnetic layer, separated by a non-magnetic metallic or insulating spacer. In these structures, the electric charges must cross the active magnetic layer to excite its magnetization dynamics. The flow of the electrical current through the active layer results in a significant Joule heating. In addition, the inhomogeneous Oersted fields induced by the localized currents can complicate the dynamical states induced by STT (see, e.g., [31,32]).

To enable current flow through the active magnetic layers, STT devices operating with spin-polarized electric current require that current-carrying electrodes are placed both on top and on the bottom of the spin valve. In addition to severely reducing the flexibility of the device geometry, this results in the optical obstruction of the active device area, making it extremely challenging to perform spatially resolved measurements of current-induced magnetization dynamics by the advanced imaging techniques [16-21,33].

An alternative approach to the implementation of STT devices that avoids these shortcomings utilizes pure spin currents – flows of spin not accompanied by directional transfer of electrical charge. This approach does not require the flow of electrical current through the active magnetic layer, resulting in reduced Joule heating and electromigration effects. One can also eliminate the electrical leads attached to the magnetic layer to drain the electrical current, enabling novel geometries and functionalities of the STT devices, as well as imaging of STT-induced dynamics by the advanced optical techniques that can provide an extraordinary insight into the underlying physical mechanisms. Moreover, it becomes possible to use insulating magnetic materials such as Yttrium Iron Garnet (YIG) [34,35]. The main advantage of this material is its exceptionally low dynamic magnetic damping. Since the expected density of the driving current necessary for the current-induced auto-oscillations is proportional to damping, YIG-based STT devices can be significantly more efficient than the traditional devices based on the transition-metal ferromagnets.

Several physical mechanisms have been utilized to create pure spin currents, including the spin-Hall effect (SHE) [36-39], the nonlocal spin injection (NLSI) [40,41], the spin Seebeck effect [42,43], and the Rashba–Edelstein effect [44,45]. We will focus in this review on the first two mechanisms, whose suitability for the excitation of coherent magnetization oscillations and waves has been already experimentally demonstrated.

The spin Hall effect is generally significant in non-magnetic materials with strong spin-orbit interaction, such as Pt and Ta. An electrical current in these materials produces a spin current in the direction perpendicular to the charge flow, due to a combination of spin-orbit splitting of the band structure (intrinsic SHE), and the spin-dependence of the electron scattering on phonons and impurities (extrinsic SHE) [38,39]. When a SHE layer is brought in contact with a ferromagnetic film, the spin current flows through the interface into the ferromagnet and exerts STT on its magnetization [46]. In recent years, SHE has been demonstrated to enable efficient magnetization switching [47-50], control of domain walls [51-53], reduction of magnetic damping [46,54-64], suppression of magnetic noise [65], and excitation of high-frequency magnetization dynamics [66-78]. The ability to exert STT on ferromagnets over extended areas is a significant benefit of SHE. Indeed, when an in-plane current flows through an extended bilayer formed by a SHE material and a magnetic film, the spin current produced by SHE is injected over the entire area of the sample, which can be as large as several millimeters [46]. This feature makes SHE uniquely suited for the control of the spatial decay of propagating spin waves in the passive regime, when STT only partially compensates the natural magnetic damping [79-84]. The benefit is less significant for the generation of coherent magnetization auto-oscillations, since the spin current injection must be spatially localized to achieve a single-mode oscillation regime [66]. Multiple dynamical modes are inevitably simultaneously excited when the injection area is extended even in only one spatial dimension [73,75].

Devices utilizing SHE benefit from many advantages of pure spin currents. However, they are also not free from shortcomings. In particular, efficient spin-Hall materials such as Ta or Pt exhibit a high resistivity. As a consequence, a significant fraction of the driving current is shunted through the active magnetic layer, which must be in electrical contact with the SHE material to enable device operation. This problem is avoided in the SHE devices based on magnetic insulators, but even in this case the high resistivity of the SHE material results in a significant Joule heating of both this material and the active magnetic layer in direct thermal contact with it. The heating adversely affects the dynamical characteristics [76] and can even damage the device. Another shortcoming of the SHE devices is the increase of magnetic damping generally observed in the bilayers of active magnetic layers with the SHE materials. This increase is difficult to avoid because it is caused by the enhanced electron-magnon scattering due to the same spin-orbit interactions that enable the operation of SHE devices.

The shortcomings described above can be avoided in devices that utilize spin currents generated by the nonlocal spin-injection (NLSI) mechanism, which does not require electric current flow through highly resistive materials [40,41,85-88]. Efficient excitation of single-frequency magnetization auto-oscillations and waves by NLSI has been recently demonstrated [23, 87-90]. In contrast to the SHE-based devices, current flow through the active magnetic layer is negligible in the NLSI-based oscillators. Their dynamical characteristics are also not compromised by the detrimental effects of layers with strong spin-orbit coupling. Moreover, since the driving current in the NLSI devices flows mostly through low-resistivity layers, the Joule heating effects are minimized.

The NLSI mechanism of spin-current generation is fundamentally different from SHE, entailing substantially different geometries of the spin-current devices. Nevertheless, the effects of spin current on the magnetization dynamics exhibit close similarities in both cases. Therefore, we will first discuss the general characteristics of the interaction of pure spin currents with the dynamic magnetization, using the SHE-based systems as an example, and then will discuss the specific features and additional opportunities offered by the NSLI mechanism.

## 2. Effects of pure spin currents on the dynamic magnetization

The effect of pure spin current on the magnetization is similar to that of spin-polarized electric currents. Both can be described by the Slonczewski's STT [91] included into the Landau-Lifshitz-Gilbert equation:

$$\dot{\mathbf{M}} = -\gamma \mathbf{M} \times \mathbf{H}_{eff} + \frac{\alpha}{M_0} \mathbf{M} \times \dot{\mathbf{M}} + \frac{\beta}{M_0^2} \mathbf{M} \times (\mathbf{M} \times \hat{\mathbf{s}}), \qquad (1)$$

where γ is the gyromagnetic ratio, α is the Gilbert damping parameter, β is the strength of the spin transfer torque proportional to the spin current density, $\mathbf{H}_{eff}$ is the effective field including the external bias field, the dipolar magnetic field, and the Oersted field of the current, $M_0$ is the saturation magnetization, and $\hat{\mathbf{s}}$ is the unit vector in the direction of the spin-current polarization. Within this model, the effect of spin current can be analyzed as a simple modification of the effective magnetic damping [46]. Correspondingly, one expects that spin current with an appropriate polarization can reduce magnetic damping, and ultimately completely compensate it when its magnitude reaches a certain critical value. Experimentally, the reduction of damping with increasing spin current can be observed as a gradual decrease of the width of the ferromagnetic resonance (FMR) curve, or alternatively as an increase in the relaxation time of magnetization precession. Complete compensation of damping is expected to result in the onset of steady-state magnetization auto-oscillations.

As will be discussed below, this qualitative picture does not take into account several important contributions to the dynamics of magnetic materials, limiting its applicability to real systems. Most significantly, it neglects the effects of fluctuations always present at finite temperatures. The fluctuations can be incorporated into Eq. (1) as an additional fluctuating random field, whose magnitude is determined by the fluctuation-dissipation relation. Analysis taking into account this contribution reveals another important effect of spin current – it drives the spin system out of thermal equilibrium, resulting in the enhancement or suppression of magnetic fluctuations [65,92]. In many cases, the influence of small magnetic fluctuations can be neglected. However, the fluctuations enhanced by the spin current often play a significant role. For instance, it was experimentally shown [65,93,94] that, in the vicinity of the damping compensation point, the intensity of fluctuations enhanced by the spin current can be up to two orders of magnitude above its equilibrium room-temperature level. As a result, magnetic fluctuations dominate the dynamical behaviors at large spin currents, and strongly influence the characteristics of current-induced auto-oscillations.

*2.1. Modulation of the effective magnetic damping*

The first demonstration of the possibility to control magnetic damping by spin current generated by SHE was reported by Ando et al. [46]. This work achieved variations of the damping constant of up to a few percent. Thereafter, the possibility to control damping by SHE in metallic and insulating magnetic films has received a significant attention [54-64]. The current-induced damping variations reported by different groups ranged from a few percent to complete compensation, in the latter case resulting in the onset of magnetic auto-oscillations [66,67].

In addition to the excitation of auto-oscillations, the ability to modify magnetic damping by SHE can be utilized for electronic tuning of the quality factor in microwave magnetic nano-resonators. This application requires not only efficient control of damping by the spin current, but also sufficiently small damping when no current is applied. Unfortunately, it is difficult to simultaneously satisfy both of these requirements in the SHE-based systems, since the same spin-orbit effects that produce spin currents also adversely affect the dynamic characteristics of the adjacent ferromagnet. As a result, the magnetic damping constant in thin ferromagnetic films in contact with the SHE materials often increases by more than a factor of two [46,54].

An approach allowing one to simultaneously achieve large variations of the damping rate and minimize the increase of magnetic relaxation associated with the proximity of the SHE material was demonstrated in [58]. The test devices used in these studies consist of a 5 nm thick and 2 μm in diameter $Ni_{80}Fe_{20}$ =Permalloy (Py) disk fabricated on top of a 10 nm thick and 2.4 μm wide Pt microstrip, as shown schematically in Figure 1(a). To reduce the adverse effects of strong spin-orbit coupling in Pt on the magnetization in Py, the Py disk is separated from the Pt microstrip by a 2 nm thick Cu spacer. The operation of the device is illustrated in the inset in Fig. 1(a). Electrons with opposite directions of the magnetic moment are scattered toward the opposite boundaries of the Pt film, due to the spin Hall effect in Pt. As a result, pure spin current is injected through the Cu spacer into the Py film, producing a spin-transfer torque *T* on its magnetization *M*. Spin torque either reduces or enhances the effective magnetic damping in the Py disk, depending on the direction of the current relative to the magnetization. The latter is aligned with the saturating static magnetic field. Damping is characterized by measurements of the ferromagnetic resonance (FMR) curves. The FMR is excited by applying a small-amplitude microwave-frequency current in addition to the dc current, producing an rf magnetic field $h$ oriented perpendicular to axis of the Pt strip. The efficiency of STT is optimized when the static magnetic field is transverse to the Pt strip. However, in this experiment the in-plane field is tilted by 10° with respect to this direction to enable the coupling of the dynamic field with the magnetization in Py. Both dc and microwave currents are either continuous, or applied in 100 ns-long pulses with a repetition period of 2 μs to reduce Joule heating.

The FMR is detected by the micro-focus Brillouin light scattering (BLS) spectroscopy technique [28,95,96]. The probing laser light generated by a single-frequency laser is focused by a 100x microscope objective lens into a diffraction-limited spot on the surface of the Py disk (Fig. 1(a)). The interaction of light with the magnetic excitations in Py results in its phase and/or amplitude modulation at the frequency of the excitations, which is analyzed by a six-pass Fabry-Perot interferometer. The resulting signal – the BLS intensity – is proportional to the intensity of the magnetic oscillations at the position of the probing spot.

Figure 1(b) shows the FMR curves recorded by BLS at different values of dc current $I$. The data were obtained by varying the frequency of the microwave signal from 7 to 10 GHz while simultaneously recording the BLS intensity at the same frequency. As seen from Fig. 1(b), at negative dc current the peak broadens and its amplitude is reduced, while at positive current is becomes narrower and its amplitude is increased. By fitting the recorded spectra with the Lorentzian function, one can determine the FMR frequency $f_0$ and the spectral width $\Delta f$ of the FMR peak. The dependence of $f_0$ on current is shown in Fig. 1(c). The results obtained with a continuous current are shown with open symbols, whereas pulsed-current results are shown with solid symbols. In both cases, the FMR frequency exhibits a significant dependence on current, due to the combined effects of Joule heating, STT, and the Oersted field $\mathbf{H}_I$ produced by the current. The latter can be calculated based on the geometrical parameters of the Pt strip. The dashed line in Fig. 1(c) shows the FMR frequency calculated using the Kittel formula [97]

$$f_0 = \gamma\sqrt{H(H + 4\pi M_0)}, \qquad (2)$$

where $\gamma$ is the gyromagnetic ratio, $4\pi M_0$ is the saturation magnetization, and $H=|\mathbf{H}_0+\mathbf{H}_I|$. The experimental data exhibit a linear variation at small $I$, closely following the calculated line. This indicates that the measured variations of $f_0$ at small $I$ are dominated by the Oersted field. At larger $I$, the measured dependencies deviate from the calculation, which can be attributed to the variations of the effective magnetization $M_e$ caused by the combined effects of STT and Joule heating. The contribution of Joule heating can be separated from that of STT by comparing the results obtained in the constant-current and the pulsed-current regime. As seen from the data of Fig. 1(c), heating significantly affects the FMR frequency only at large currents.

Based on the measured values of $f_0$ and $\Delta f$, one can determine the current-dependent effective magnetization $M_e(I)$ and the effective Gilbert damping parameter $\alpha_{\text{eff}}(I)$ of the Py disk. The value of $M_e(I)$ can be calculated using the Kittel formula Eq. (2). One can determine $\alpha_{\text{eff}}(I)$, from the expression

$$\alpha_{\text{eff}} = \frac{\Delta f}{2\gamma(H+2\pi M_e)} \qquad (3)$$

derived from the Landau-Lifshitz-Gilbert (LLG) equation, taking into account the demagnetization effects for an in-plane magnetized film [98]. The results of calculations are shown in Fig. 1(d). The value $\alpha_{\text{eff}}=0.011$ at $I=0$ is close to the standard value $\alpha=0.008$ for Permalloy, demonstrating that the effect of Pt on damping in Py is minimized by the insertion of the Cu spacer. The value of $\alpha_{\text{eff}}$ varies by about a factor of 4 within the studied range of current, and the smallest achieved value $\alpha_{\text{eff}}=0.004$ is only half of the standard value for Permalloy, demonstrating that SHE can be utilized as an efficient and practical mechanism for controlling the dynamical characteristics of ferromagnets. Moreover, the obtained dependence $\alpha_{\text{eff}}(I)$ [Fig. 3(b)] is the same for the continuous and the pulsed current, indicating that damping is not significantly affected by Joule heating.

In addition to the FMR measurements, the variation of the effective damping due to SHE can be determined from the time-resolved measurements of the temporal decay of magnetization precession excited by pulsed current [62], or by analyzing the parametric instability threshold [99,100]. Both of these methods yield a linear dependence of the effective damping constant on current, consistent with the FMR measurement discussed above. We emphasize that a simple linear dependence is observed only at modest currents. More complex dynamical behaviors of the magnetic system are expected at spin currents approaching the point of complete damping compensation, due to the increasingly significant role of nonlinear phenomena. As will be discussed in the next Section, these deviations are caused by the strong enhancement of magnetic fluctuations concurrent with the reduction of effective damping.

*2.2. Enhancement and suppression of magnetic fluctuations*

The effects of STT on magnetic fluctuations have been extensively investigated for traditional devices driven by the spin-polarized electric current. For example, studies of magnetization reversal in nano-elements showed that STT can modify their thermal activation rates, which was interpreted as evidence for the effect of STT on thermal fluctuations [101,102]. This effect was measured directly by noise spectroscopy in magnetic tunnel junctions, taking advantage of their large magnetoresistance [103-105]. However, such electronic measurements require a finite dc bias and are limited to magnetic configurations producing large magnetoresistive signals. In contrast to the typical multilayer STT devices that require nontransparent electrical contacts to the magnetic layers, the geometry of devices driven by pure spin current allows optical access to the active magnetic layer. Therefore, these systems can be studied by the BLS spectroscopy, which is uniquely suited for the detection of magnetic fluctuations due it its unmatched sensitivity. In particular, the BLS spectroscopy is capable of detecting magnetic fluctuations, always present at room temperature, even in the absence of electrical current.

The experiments described below [65] were performed with a sample similar to that shown in Fig. 1(a), but omitting the Cu spacer between the Py and Pt films. Since these measurements do not require excitation of magnetization dynamics by the high-frequency magnetic field, only dc current was applied, and the static magnetic field $H_0$ was oriented perpendicular to the Pt microstrip to maximize the efficiency of SHE. Magnetic fluctuations were detected by BLS, which yielded a signal proportional to the square of the dynamic magnetization, or equivalently to the energy associated with the magnetic fluctuations. In the FMR measurements, the quasi-uniform precession mode of the sample driven by the rf field dominates over all other modes. In contrast, in the fluctuation spectroscopy measurements, all the magnetic modes are simultaneously excited due the interactions between the spin system and the lattice. In thermal equilibrium, the intensities of the modes follow the Bose-Einstein statistics with zero chemical potential [106,107].

The spectrum of the spin-wave modes [108,109] is schematically shown in the right panel of Fig. 2(a). The two solid curves show the dispersions of the spin waves propagating perpendicular and parallel to the direction of the static magnetic field, as labeled. These curves are the boundaries of the continuous spin-wave manifold shown by the shaded area. At finite temperatures, all these modes contribute to the magnetization fluctuations. The BLS microscopy is capable of detecting spin-wave modes within a limited range of wavevectors 0-$k_{max}$~$10^5$ cm$^{-1}$ [95,96], and its sensitivity continuously decreases with the increase of the mode wavevector. As a result, the BLS spectrum of magnetic fluctuations is typically shaped as an asymmetric peak with the maximum at the frequency $f_0$ of the FMR (Fig. 2(a) left panel). The amplitude of this peak characterizes the intensity of the long-wavelength magnetic modes, while the frequency $f_0$ can be used to determine the effective static magnetization of the sample.

Figure 2(b) shows representative spectra of magnetic fluctuations recorded at three different values of dc current applied to the Pt microstrip, $I$=–26 mA, 0, and 26 mA. The magnetic moments in the spin current are parallel to the magnetization at $I$<0, and antiparallel to it at $I$>0. As discussed above, in the former case the SHE-induced STT is expected to enhance the effective magnetic damping, consistent with the decrease of the fluctuation peak's amplitude at $I$<0. Similarly, the increase of the fluctuation peak amplitude at $I$>0 is consistent with the expected decrease of the effective damping. However, the modification of damping cannot account for the dependence of the integral intensity of the BLS spectra on current, shown with point-up triangles in Fig. 2(c). The integral intensity of the BLS peak is proportional to the average fluctuation energy of the long-wavelength magnetic modes. In the classical limit applicable to these modes, the equilibrium fluctuation energy associated with each dynamical mode is $k_BT$, regardless of damping [110]. If the magnetic system were to behave simply as if the damping were changed while maintaining thermal equilibrium, the integral intensity would remain constant. In contrast, at $I$=28 mA the integral intensity increases relative to the equilibrium value by more than a factor of 30, while at $I$=-28 mA it decreases by more than a factor of two. These results clearly demonstrate that in addition to modifying the damping, STT brings the magnetic system out of equilibrium. The observed behaviours are consistent with the established theories of STT, once different contributions to the dissipation and the associated fluctuating fields are separately considered [92,65]. The theory predicts a linear dependence of the inverse integral intensity on current, in agreement with the data shown in Fig. 2(c) by the point-down triangles. One can extrapolate the low-current linear variations of the inverse intensity to determine the intercept $I_C$ =28 mA, corresponding to the critical current at which the intensity of the BLS peak diverges due to the complete damping compensation. At currents above $I_C$, the system can be expected to enter the auto-oscillation regime [92]. Instead, the integral intensity saturates and starts to decrease at $I>26$ mA (Fig. 2(c)), indicating an onset of a new relaxation process that limits the amplitude of magnetic fluctuations and prevents the onset of current-induced auto-oscillations.

The central frequency $f_0$ of the peak exhibits a red shift at $I>0$ (point-up triangles in Fig. 2(d)) that becomes particularly significant at large currents. It is caused by the decrease of the effective magnetization $M_e$, due to the increased intensity of magnetic fluctuations. The magnitude of $M_e$ can be determined from the measured $f_0$ by using the Kittel formula (Eq. (2)), once the Oersted field of the dc current is incorporated into the expression for the magnetic field $H$. The magnitude of $M_e$ (point-down triangles in Fig. 2(d)) monotonically decreases with increasing current even at $I>26$ mA. Thus, the total intensity of magnetic fluctuations keeps increasing at large currents, even after the intensity of long-wavelength fluctuations saturates and starts to decrease. This can be explained by the different behaviours of different spin wave modes. The BLS spectra are selectively sensitive to the long-wavelength fluctuations, while the total fluctuation intensity is dominated by the much larger phase volume of short-wavelength modes. Thus, as the current approaches the critical value, only the long-wavelength fluctuations become suppressed resulting in the saturation of the BLS intensity, while the short-wavelength fluctuations continue to increase resulting in decreasing $M_e$.

Additional information about the mechanisms contributing to the phenomena observed at large currents was obtained from time-resolved BLS measurements, performed with dc current applied in 1μs long pulses. Figure 2(e) shows the temporal evolution of the integral BLS peak intensity (point-up triangles) and of the effective magnetization $M_e$ (point-down triangles) obtained at current $I$=25 mA below the onset of the saturation of the fluctuation intensity. Figure 2(f) shows the results of the same measurements for $I$=30 mA, above the onset of saturation. In both cases, the value of $M_e$ rapidly changes at the onset of the pulse, and subsequently varies on a much longer characteristic timescale. At the end of the pulse, $M_e$ first rapidly increases, and then slowly relaxes. These results allow one to separate the contribution of STT from the Joule heating. The rapid increase of $M_e$ at the end of the pulse can be attributed to the relaxation of the magnetic system towards equilibrium with the lattice. This process is characterized by the spin-lattice relaxation rate of a few nanoseconds. The subsequent slow relaxation of $M_e$ is associated with the simultaneous cooling of the lattice and the magnetic system. Therefore, by comparing the magnitudes of the fast and the slow variations of $M_e$ at the end of the pulse, one can estimate that the relative contribution of Joule heating to the total enhancement of fluctuations does not exceed 30%.

While the temporal evolution of $M_e$ is similar both at $I$=25 mA (Fig. 2(d)) and at $I$=30 mA (Fig. 2(e)), the evolution of the BLS intensity is qualitatively different in these two cases. At $I$=25 mA, the intensity abruptly jumps at the onset of the pulse, and subsequently continues to slowly increase. At $I$=30 mA, the intensity also initially abruptly increases, but then slowly decreases over the rest of the pulse duration. These different temporal behaviours likely originate from the same nonlinear dynamical mechanisms that lead to the saturation of intensity in the static measurements (Fig. 2(c)). Since the initial increase of intensity at $I$=30 mA is significantly larger than at $I$=25 mA, it is the subsequent slow variation that results in the saturation seen in the static measurements. By examining the temporal evolution of $M_e$ and the intensity, one can infer that the fluctuations of both the long- and the short-wavelength modes are initially significantly more enhanced at $I$=30 mA than at $I$=25 mA. This enhancement leads to nonlinear coupling among different modes, resulting in the redistribution of the fluctuation energy within the spin-wave spectrum. Such nonlinear mode coupling can be generally expected to drive the magnetic subsystem towards a thermalized distribution [111,112]. Since the long-wavelength fluctuations are most significantly enhanced by STT due to their lower damping, they can be also expected to be most significantly suppressed by the nonlinear scattering, preventing the onset of auto-oscillation. As the fluctuation starts to diverge when the critical current is approached, the nonlinear scattering processes that suppress long-wavelength fluctuations become increasingly efficient. Consequently, complete compensation of the magnetic damping by SHE cannot be achieved in the discussed system because of the additional damping that emerges at large currents due to the nonlinear scattering processes. While the detailed mechanisms of these processes are still unclear, one can expect that in a strongly nonequilibrium spin system characterized by the reduction of the effective magnetization by more than a factor of two (Fig. 2(d)), the nonlinear interactions can be mediated not only by the resonant three- and four-magnon scattering processes [98,113], but can also include complex non-resonant processes that do not require exact phase synchronism of the interacting modes (see, e.g., [114-116]).

The results discussed above clearly demonstrate that one of the main difficulties in the implementation of magnetic auto-oscillators driven by pure spin current is associated with the non-selective effect of STT. The spin torque simultaneously enhances all the dynamic magnetic modes, resulting in the onset of nonlinear interactions that limit the intensity of the fundamental quasi-uniform precession mode. In the next Section, we will discuss an approach allowing one to overcome this difficulty and achieve magnetization auto-oscillations in SHE systems.

## 3. Spin Hall-effect magnetic nano-oscillators based on metallic ferromagnets

One of the main benefits of spintronic and magnonic devices based on pure spin currents is their compatibility with low-loss magnetic insulators. However, the first demonstrations of coherent magnetization auto-oscillations driven by SHE were reported for all-metallic spin-current systems [66,67]. Extensive subsequent studies of these systems have produced novel device geometries and significant improvement of the oscillation characteristics [68-75,87-90]. In contrast, the insulator-based devices, which will be discussed in details in Section 4, are only starting to emerge. Their performance is still significantly lagging behind the all-metallic spin-current oscillators.

*3.1. Self-localized bullet oscillation mode in Spin-Hall systems*

As discussed in the previous Section, damping reduction by SHE can be relatively easily achieved at small current densities. However, complete compensation of damping by the spin current, which is necessary to achieve steady-state magnetization auto-oscillations, is not a straightforward extension of the damping reduction. The simultaneous enhancement of many dynamical modes by spin current results in nonlinear damping enhancement, preventing the transition to the auto-oscillation regime. Therefore, it is necessary to suppress the nonlinear interactions to achieve auto-oscillation. Since magnon-magnon scattering rates are proportional to the populations of the corresponding modes, the detrimental effects of nonlinear damping can be avoided by selectively suppressing all the modes, except for the ones that can be expected to auto-oscillate. To achieve such a selectivity, one can take advantage of the frequency-dependent damping caused by the spin-wave radiation in a system based on the local injection of spin current into an extended magnetic film [66].

The schematic of the device supporting mode-selective radiation losses is shown in Fig. 3(a). The device is formed by a bilayer of a 8 nm thick film of Pt and a 5 nm thick film of Permalloy patterned into a disk with a diameter of 4 µm. Two 150 nm thick Au electrodes with sharp points separated by a 100 nm wide gap are placed on top of the bilayer, forming an in-plane point contact. The sheet resistance of the bilayer is nearly two orders of magnitude larger than that of the Au electrodes. As a consequence, the electrical current induced in the bilayer by the voltage between the electrodes is localized predominantly within the nano-gap region. The electric current produces a pure spin current due to the spin-Hall effect in Pt. The spin current is injected into the Py film in a relatively small active area defined by the geometry of the electrodes (see the inset in Fig. 3(a)). In this area, the local spin current enhances a large number of different dynamical modes. The high-frequency spin-waves are characterized by large group velocities (see the spin-wave spectrum in Fig. 2(a)). Consequently, they quickly escape from the active region, resulting in their efficient suppression by the radiation losses. Meanwhile, the low-frequency modes with large wavelengths have a much smaller group velocity, and therefore their radiation losses are minimal. Moreover, as will be discussed below, this damping imbalance is further enhanced by the formation of a self-localized "bullet" mode [117], which is completely free from the radiation losses.

Figures 3(b) and 3(c) show the BLS spectra obtained with the probing laser spot positioned in the center of the nano-gap, at different values of the dc current $I$. At $I$=0 (Fig. 3(b)), the BLS spectrum exhibits a broad peak produced by incoherent magnetization fluctuations in the Py film. Similarly to the systems driven by spatially uniform spin currents (see Section 2.2), this peak grows with increasing current due to the enhancement of fluctuations by SHE. Additionally, its rising slope becomes increasingly sharper than the trailing slope, consistent with the selective preferential enhancement of the low-frequency modes. In contrast to the systems driven by spatially uniform spin currents, the integral BLS intensity does not saturate when the current approaches the critical value $I_C \approx 16.1$ mA. Instead, a new peak appears in the BLS spectrum at $I_C$ at a frequency below the fluctuation peak, as indicated in Fig. 3(b) by an arrow. The intensity of this peak rapidly grows with increasing current, and then saturates at a value that exceeds the intensity of thermal fluctuations by more than two orders of magnitude (Fig. 3(c)). These behaviours clearly indicate the onset of the auto-oscillations, and show that the radiation-loss mode-selection mechanism allows one to overcome the nonlinear effects that prevent auto-oscillation in systems with spatially extended spin current injection. Moreover, by comparing the spectra obtained at $I$=16.1 mA and 16.3 mA, one can infer that the onset of auto-oscillations is accompanied by a decrease in the intensity of the fluctuation peak, suggesting that the energy of spin current is mainly channelled into the auto-oscillation mode.

Since the peak corresponding to the auto-oscillation mode is not present in the thermal fluctuation spectrum, one can conclude that it represents a new mode that does not belong to the spin-wave spectrum. Since energy can be radiated only by propagating spin waves and there are no available spin-wave spectral states at the auto-oscillation frequency, the auto-oscillation mode must be completely free from the radiation losses. The nature of this new mode can be elucidated based on the two-dimensional BLS mapping of the dynamic magnetization at the frequency of auto-oscillations, which can be obtained by rastering the probing laser spot in the two lateral directions and simultaneously recording the BLS intensity. An example of the obtained maps is presented in Fig. 3(d). These data show that the auto-oscillations are localized in a very small area in the gap between the electrodes. The measured spatial distribution is a convolution of the actual spatial profile with the instrumental function determined by the shape of the laser spot. Taking into account the size of the spot of about 250 nm, one can estimate from the measured distribution that the size of the auto-oscillation region is less than 100 nm, significantly smaller than the characteristic size of the current localization. This indicates that the auto-oscillation area is determined not by the spatial localization of the spin current, but by the nonlinear dynamical properties of the magnetic system. Such nonlinear self-localization resulting in the formation of a standing spin-wave soliton (or spin-wave bullet) has been predicted for extended magnetic films subjected to local spin-current injection [117]. This conclusion is further supported by the results of numerical micromagnetic simulations [118,119], which reveal that a self-localized bullet with typical dimensions of about 80 nm is formed in the studied system.

The planar nano-gap oscillator is robust with respect to the experimental parameter variations. Figure 3(e) shows the static-field dependences of its oscillation characteristics. As seen from these data, the devices exhibit auto-oscillations for all static fields in the interval from 400 to 2000 Oe, and their oscillation frequency can be efficiently tuned by the field within a broad range. The auto-oscillation frequency (point-up triangles in Fig. 3(e)) always remains below the FMR frequency (point-down triangles) in the Py film, as expected for the nonlinear self-localized bullet mode. The oscillation onset current (diamonds) exhibits a modest variation of about 5% within the entire field range.

Because of the limited spectral resolution of the BLS spectroscopy, the measured BLS spectra (Fig. 3(c)) cannot provide accurate information about the spectral width of the auto-oscillation peak, which is an important parameter characterizing the coherence of the spin-current induced oscillations. Measurements of the oscillation linewidth of the nano-gap oscillators were performed by electronic microwave spectroscopy in [68]. In particular, it was shown that these devices produce a relatively large microwave power and exhibit a small auto-oscillation linewidth at cryogenic temperatures. However, both of these characteristics significantly degrade as the temperature is increased. This result is not surprising, since the small size of the self-localized bullet mode leads to a small total oscillation energy comparable to the energy of thermal fluctuations at room temperature. As a result, the coherence of the auto-oscillations is strongly affected by the thermal noise. In principle, these effects can be avoided by modifying the geometry of the oscillator. For example, a moderate improvement of the spectral characteristics of planar point-contact oscillators was achieved by patterning of the Pt layer into a 400 nm diameter disk centered in the active device region [71]. This improvement was likely caused by the more efficient radiation of spin waves from the active device region, since the area of Py interfaced with Pt, which is detrimental to the magnetization dynamics, was reduced in these devices. Overall, the possibility to control the auto-oscillation characteristics by modifying the device geometry is limited, since local injection of spin current into a continuous magnetic film leads to the spontaneous formation of the bullet mode, whose spatial dimensions are determined by the nonlinear self-localization effects rather than by the spin-current injection area.

*3.2. Effects of confining effective potential in Spin-Hall oscillators*

Spin-Hall nano-oscillators capable of efficient room-temperature generation of coherent microwave signals were demonstrated in [72]. The improvement was achieved by controlling the auto-oscillation characteristics via magnetic dipolar effects instead of the self-localization. Specifically, the localized auto-oscillation mode was formed due to the confinement in an effective potential well produced by the non-uniformity of the internal static magnetic field in a bow tie-shaped magnetic nanoconstriction. The devices based on this principle are characterized by a large oscillation area, minimizing the effects of thermal fluctuations and resulting in a narrow room-temperature spectral linewidth without compromising the single-mode regime of auto-oscillation.

A schematic of these devices is shown in Fig. 4(a). They are fabricated from a 4 µm wide Py(5 nm)Pt(8 nm) bilayer strip. A sharp bow tie-shaped nanoconstriction with the width of 150 nm, radius of curvature of about 50 nm, and an opening angle of $22^o$ is defined in the center of this strip. The principle of the device operation is very similar to the nano-gap oscillators discussed in Section 3.1. The electrical current flowing in the Pt layer induces a spin current due to SHE, which is injected into Py. The abrupt narrowing of the current-carrying Pt layer in the nanoconstriction results in a strong increase of the local current density, which exhibits a sharp peak with the width of about 0.3 µm along the device axis (see the calculated current distribution in the inset in Fig. 4(a)). Since the spin current injected into Py is proportional to the current density in Pt, this region of large current density also defines the active device area, in which the injected spin current is sufficiently large to cause a sizable effect on the magnetization. Note that the active region is surrounded by the magnetic film that supports propagating spin waves, similarly to the nano-gap devices. Consequently, the mode-dependent radiation-loss mechanism that suppresses the nonlinear mode coupling phenomena is also effective in the nanoconstriction devices.

The symmetry of SHE implies that the largest effect of spin current is achieved when the magnetization is oriented perpendicular to the direction of the electric current flow. However, if the applied static field $H_0$ is exactly perpendicular to the direction of the current, the current-induced oscillations cannot generate electronic signals at the oscillation frequency due to the anisotropic magnetoresistance (AMR) effect. Therefore, to enable electronic detection of auto-oscillations, in the experiments described below the field was applied in-plane at an angle $\alpha=60°$ relative to the current (Fig. 4(a)).

Figures 4(b) and 4(c) show the results of device characterization by the electronic microwave spectroscopy. A dc current $I$ was applied to the nanoconstriction through the low-frequency branch of a microwave bias tee. The resulting current-induced magnetization oscillations were converted by the AMR effect into a microwave voltage, which was detected by a spectrum analyzer. A narrow spectral peak appeared in the microwave spectra at $I>I_C\approx3.3$ mA (Fig. 4(b)). The intensity of the detected peak gradually increases with increasing current while its spectral linewidth decreases (Fig. 4(c)), reaching a minimum of 6.2 MHz at $I=3.75$ mA. The integral intensity of the microwave signal reaches a maximum of about 7 pW at $I=3.95$ mA. The power generated by this device is comparable to that delivered by the giant-magnetoresistance spin-torque oscillators operating at comparable frequencies (see, e.g., [14] and references therein), despite the relatively low magnitude of the AMR effect. At $I>3.95$ mA, the generation power abruptly decreases, while the auto-oscillation peak broadens and splits into multiple peaks, marking a transition to the multimode auto-oscillation regime.

The large generated microwave power can be attributed to the large oscillation amplitude of the nanoconstriction region that provides a dominant contribution to the device resistance, while the small generated linewidth is likely associated with the large volume of the oscillating magnetization, making the oscillator insensitive to thermal fluctuations. This hypothesis was confirmed by micro-focus BLS measurements. Figure 4(d) shows the BLS spectra of the magnetization oscillations recorded at different currents. The BLS spectrum acquired at $I=0$ reflects the thermal magnetization fluctuations. At small $I$ (1 and 3 mA in Fig. 4(d)), the fluctuations are enhanced and a peak appears on the low-frequency shoulder of the thermal fluctuation background. This peak rapidly increases in intensity at $I>I_C$ (3.8 mA in Fig. 4(d)), marking the onset of current-induced magnetization auto-oscillations. Note that the spectral peak corresponding to the auto-oscillation mode does not abruptly appear at the onset current, as was the case for the nano-gap oscillators, but instead gradually emerges at $I<I_C$. This behavior can be contrasted with the spin-wave bullet observed in the nano-gap oscillator, which formed spontaneously at the oscillation onset due to the nonlinearity associated with the large amplitude of auto-oscillation. Further evidence against the formation of the "bullet" mode in the nanoconstriction devices is provided by the smooth variation of the center frequency of the observed spectral peak (dashed line in Fig. 4(d)). In contrast, the formation of the bullet mode is expected to be accompanied by a frequency jump.

Spatially resolved BLS measurements (left panel in Fig. 4(e)) further elucidated the nature of the oscillation mode. As seen from Fig. 4(e), the oscillation mode is localized in the nano-constriction area, and exhibits an elliptical shape elongated in the direction perpendicular to the external field $H_0$. Taking into account the finite size of the probing laser spot, one can estimate from the BLS data that the spatial dimensions of the localization area are 0.25 μm in the direction parallel to $H_0$, and 0.4 μm perpendicular to it. These values are significantly larger than the typical spatial dimensions of about 80 nm expected for the bullet mode (Section 3.1). The nature of the auto-oscillation mode was clarified by micromagnetic simulations of the static magnetic state of the nanoconstriction. The right panel in Fig. 4(e) shows the spatial distribution of the demagnetizing field in the area of the nanoconstriction, calculated using the OOMMF micromagnetic simulation software [120]. The calculations show that the edges of the nanoconstriction produce a strong demagnetizing field that opposes the external static magnetic field, resulting in a local decrease of the internal field in Py in the nanoconstriction area. The local reduction of the static internal field creates an effective potential well for spin waves [121,122] resulting in the formation of a localized dynamic mode. This mechanism is similar to the well-known effect of quantum state localization of particles in potential wells. In contrast to all the other dynamical modes of the system, the localized mode is free from radiation losses. Consequently, it is preferentially enhanced by the spin current, resulting in single-mode auto-oscillations. This auto-oscillation mechanism is qualitatively different from that in the nano-gap oscillators, where only propagating linear modes exist in the linear regime at small currents, and the localized bullet mode is spontaneously formed only at the onset of auto-oscillation due to the nonlinear self-localization effects.

While the operational characteristics of SHE oscillators can be significantly improved by increasing the spatial size of the auto-oscillation mode using the dipolar-field localization mechanism, this increase inevitably leads to a multimode auto-oscillation regime observed in nanoconstriction devices at large currents (Fig. 4(c)). The oscillation characteristics significantly degrade in this regime, likely due to thermal mode hopping. Therefore, this regime is not beneficial for technical applications of the SHE oscillators. Note that further increase in the spatial size of the auto-oscillation area up to 2 μm, which was implemented in the nano-wire spin-Hall oscillators [73], does not result in improvement of the oscillation characteristics. Instead, single-mode auto-oscillations are observed only in a very narrow range of parameters. It was recently shown that the single-mode oscillation regime is more robust in a tapered nanowire [75]. This improvement is mostly caused by the reduction of the auto-oscillation area due to the non-uniform distribution of the current density along the tapered nanowire. The limiting case of the tapered-nanowire geometry is the nanoconstriction oscillator discussed above.

Since the increase in the dimensions of the auto-oscillation region does not allow an improvement in the microwave power generated by the SHE devices, the most straightforward route to achieve such an improvement is associated with the implementation of arrays of mutually synchronized oscillators, as was previously suggested for traditional STT devices [22,123-126].

*3.3 Synchronization of spin-Hall oscillators*

One of the distinguishing characteristics of STT-driven magnetic nano-oscillators is a strong nonlinearity that enables their efficient synchronization to external periodic signals over a wide frequency range, as well as mutual synchronization of oscillators with significantly different frequencies [92]. This feature may allow the development of microwave sources with improved generation coherence using mutually synchronized oscillator arrays. The implementation of such arrays with spin-Hall oscillators is significantly more straightforward than with the conventional ones due to their simpler planar geometry.

The ability of SHE-based oscillators to synchronize to external signals was first demonstrated in [69]. In this work, the characteristics of synchronization were studied for nano-gap devices discussed in detail in Section 3.1. To study the effects of external signals on the oscillation characteristics, a microwave current was applied simultaneously with the dc current $I$. The microwave current induced a dynamic Oersted magnetic field in the Py film. Since the operation of spin-Hall oscillators is optimized when the static magnetic field is perpendicular to the current flow, such that the dynamic field created by the microwave current is parallel to the static magnetization, the dynamic magnetic field can only couple to the dynamic magnetization through the parametric process that becomes efficient when the driving frequency is close to twice the auto-oscillation frequency [98]. Under these conditions, the oscillators were found to exhibit efficient parametric synchronization characterized by a relatively wide synchronization frequency interval controlled by the magnitude of the microwave signal and by the dc current flowing through the device. In contrast to the conventional STT oscillators, the synchronization of spin-Hall oscillators occurred only above a certain threshold microwave signal power, which was attributed to the influence of magnetic fluctuations enhanced by spin-current, whose importance was discussed in Section 2.2.

Figure 5(a) shows the BLS spectra recorded at $I$=16.5 mA, in the stable auto-oscillation regime of the nano-gap device with a saturated intensity (see also Fig. 3(c)). The data were recorded with the frequency $f_{MW}$ of the microwave signal varying from 12 to 16 GHz and the power fixed at $P$=0.5 mW. The spectra exhibit a narrow intense auto-oscillation peak, regardless of the external microwave frequency. When $f_{MW}/2$ approaches the auto-oscillation frequency, the latter starts to exactly follow $f_{MW}/2$ (dashed line in Fig. 5(a)), as expected for the parametrically synchronized oscillation. In addition to the frequency locking demonstrated by the BLS measurements, the reduction of the auto-oscillation linewidth in the synchronized regime was analyzed by the electronic spectroscopy. These measurements demonstrated that at cryogenic temperatures the linewidth of the auto-oscillation in the synchronized regime is reduced by up to four orders of magnitude (Fig. 5(b)).

Figure 5(c) shows the dependence of the synchronization interval $\Delta f_S$ (see Fig. 5(a)) on the dynamic microwave magnetic field, which is proportional to the square root of the driving microwave power $P$. No synchronization was observed at microwave powers below $P^S_{th} \approx 0.03$ mW, regardless of the dc current value. At a current $I$=18 mA significantly above the oscillation onset, $\Delta f_S$ exhibits an approximately linear dependence on $P^{1/2}$ above the threshold $P^S_{th}$ (diamonds and solid straight line in Fig. 5(c)), while at smaller currents close to the oscillation onset, a rapid increase of $\Delta f_S$ at small microwave power is followed by saturation (triangles in Fig. 5(c)). Note that the observed behaviors cannot be described by the theory of synchronization of nonlinear single-mode oscillators [92], which predicts a universal linear dependence of $\Delta f_S$ on $P^{1/2}$. The dependence predicted by this theory also starts at the origin, since the synchronization is known to be a non-threshold phenomenon. Theoretical analysis performed in [69] showed that this inconsistency can be explained by the effects of the magnetic fluctuations enhanced by the spin current. In particular, by including the effects of fluctuations in the auto-oscillator model, the non-zero power threshold of the synchronization process was obtained. The effective amplitude of fluctuations, estimated by comparing the theoretical and experimental results, is about two orders of magnitude larger than its thermal-equilibrium magnitude at room temperature. This result is consistent with the expected influence of spin current, which brings the magnetic system into a strongly nonequilibrium state (Section 2.2). Moreover, it was found that the synchronization threshold was not reduced even at cryogenic temperatures, indicating that the magnetic noise that determines the threshold is dominated not by the thermal fluctuations of the lattice, but rather by the magnetic modes enhanced by the spin current.

The results discussed above demonstrate the ability of SHE-driven oscillators to synchronize to external signals, which is a necessary, although not sufficient, condition for the oscillators' ability to mutually synchronize. Note that the layout of the nanoconstriction spin-Hall oscillators is uniquely suited for the studies of mutual synchronization. Indeed, it enables easy implementation of the oscillator chains, by using a Py/Pt strip with a periodic series of nanoconstrictions. The individual oscillators in the chain can phase-lock to each other due to the exchange interaction, if their oscillation areas slightly overlap. Because of the relatively large oscillation area, such overlap can be easily achieved by placing the neighboring nanoconstrictions at a distance of 200-300 nm from each other. This separation allows a sufficiently large modulation of the driving current density along the device axis, resulting in well-defined separate effective potential wells defining the neighboring oscillators. An additional advantage provided by this layout is that the current required for the oscillations in a chain is independent of the number of oscillators. The possibility of mutual synchronization in chains of nanoconstriction devices was demonstrated in numerical simulations [127] and, very recently, was confirmed experimentally [78].

**4. Spin Hall-effect magnetic nano-oscillators based on magnetic insulators**

We now turn our attention to the study of injecting pure spin current into magnetic insulators. The microscopic mechanisms of transfer of angular momentum between a normal metal and a ferromagnetic layer are now quite different. While in the previous case of a metallic ferromagnet, electrons in each layer have the possibility to penetrate the other one, for magnetic insulators the transfer takes place exactly and solely at the interface. As mentioned in the introduction, insulating ferromagnets (garnets, ferrites) benefit from a much lower damping parameter than metals. This is because the disappearance of the electronic density at the Fermi level permits to get rid of a potent relaxation mechanism: the electron-magnon scattering [128].

Among different magnetic insulators, the material which has the lowest known damping is Yttrium Iron Garnet ($Y_3Fe_5O_{12}$, YIG) [129]. Quite importantly YIG can be synthesized in large volume in the form of a single crystal with almost no atomic disorder (FMR linewidths in polycrystals are usually much larger than those in single crystals). The smallest linewidths have been measured on polished YIG spheres of millimeter size. Three principal magnetic relaxation channels have been identified in these YIG spheres [129,130]. The potentially strongest one is linked to the exchange interaction. YIG is a ferrimagnet, which has a compensated magnetic moment on the $Fe^{2+}$ and $Fe^{3+}$ sites, both coupled by super-exchange. The associated relaxation process (the valence exchange relaxation) involves a charge transfer between the two ions, which can be activated by impurities. This effect is minimized when the YIG crystal is grown from ultra-pure material composition. At liquid nitrogen temperature, trace amount of impurities would normally dominate the relaxation. It appears in the form of an enhancement of the damping in a narrow temperature window, where the fluctuation rate of the charge transfer matches the Larmor frequency. The second potential relaxation channel is linked to the dipolar interaction. The associated relaxation process is the two-magnon decay of the uniform mode into degenerate modes due to pit scattering at the sample surface. This effect is minimized by polishing the sphere to perfection. For sample size of the order of the magnon propagation length, this broadening is inversely proportional to the sphere diameter. The third potential relaxation channel is due to spin-orbit coupling, which also sets the magnetic anisotropy. In the case of YIG this coupling is relatively small. It leads to a magnon-phonon coupling described by Kasuya and LeCraw [131], which can become the dominant relaxation channel at room temperature. This last contribution determines the ultimate linewidth of YIG. The resulting broadening at the center of the X-band (10 GHz) is about 0.25 Oe (or 0.7 MHz) at room temperature (corresponding to $\alpha = 3\times10^{-5}$), and possibly 0.01 Oe at liquid Helium temperature.

The situation, however, is very different for YIG thin films. These films are usually grown on Gadolinium Gallium Garnet, $Gd_3Ga_5O_{12}$ (GGG) substrate, which provides the necessary lattice matching to achieve epitaxial growth. Figure 6(a) shows the FMR spectrum measured at 6 GHz for a 20 nm thick YIG thin film grown by pulsed laser deposition (PLD) [132]. The 1.9 Oe separation between the two extrema of the derivative of the FMR absorption peak corresponds to the linewidth of 3.3 Oe. This linewidth is determined by a homogeneous and an inhomogeneous contributions, which can be separated by studying its frequency dependence. The Gilbert damping coefficient is extracted from the slope of a linear fit (see solid line Fig. 6(b)), and is found to be about $\alpha=2\times10^{-4}$. The finite ordinate intercept, $\Delta H_0=2.4$ Oe, represents the inhomogeneous part of the linewidth. Still the homogeneous part of the linewidth (1.6 Oe at 10 GHz) is almost an order of magnitude larger than the one reported previously for an isolated YIG sphere. The dominant relaxation channel in this case is the dipole-dipole interaction with the paramagnetic substrate [133]. Although the spin susceptibility of the GGG is reduced at room temperature and the FMR mode probed in Fig. 6(a)-(b) is significantly shifted from the paramagnetic resonance of the GGG substrate, the remaining fluctuations are still sufficient to completely dominate all the other relaxation channels. This effect only gets worse at liquid nitrogen or liquid Helium temperature because the paramagnetic spin susceptibility of the substrate increases inversely proportional to the temperature.

The damping coefficient can also be inferred from the magnon decay length measured by propagating spin-wave spectroscopy [134]. Here two coplanar waveguides with different separation between them are patterned on top of the 20 nm thick YIG. The signal attenuation between the two antennas is then measured by a vector network analyzer. The characteristic propagation length of magnons depends strongly on their wave-vector. One observes that the characteristic propagation length of the longest wave-vector magnons lies in the range of few hundreds micrometers [134].

High quality level of YIG ultra thin films can also now be achieved by using liquid phase epitaxy (LPE), the reference method to grow micrometer thick YIG [135]. Recently, this growth method has allowed to obtain variable thickness YIG films from 200 nm down to 20 nm. The measured values of the Gilbert damping coefficient of these films are shown by the blue symbols in Fig. 7(a). For the lowest LPE YIG thickness (20 nm), a linear fit of the frequency dependence of the linewidth yields $\alpha=4\times10^{-4}$ and $\Delta H_0=1$ Oe. The combined low values of *both* $\alpha$ and $\Delta H_0$ are believed to be the minimum reported in the literature for ultrathin YIG film. When a few nanometers (7 nm) thick Pt layer is added on top of these YIG films, one observes an additional increase of the FMR linewidth [136], which varies inversely proportional to the YIG thickness (see red dots Fig. 7(a)). This indicates that a new relaxation channel has been opened up where angular momentum can escape through the interface and get absorbed in the metal through the so-called spin pumping effect [137]. Note that even for YIG, whose natural linewidth is only a few Oersted at 10 GHz, the additional broadening produced by the Pt is hardly observable if the thickness of the magnetic film, $t_{YIG}$, exceeds a few hundreds of nanometers. By itself, the enhancement of the damping reported in Fig. 7(a) is an unambiguous evidence that spin current is transmitted at the interface between YIG and Pt. Furthermore, the hyperbolic dependence highlighted by the dashed curve is the proof that this new relaxation is a purely interfacial effect [138,139]. The spin transmission at the interface is parametrized by $g_{\uparrow\downarrow}$, the spin mixing conductance normalized by the quantum of conductance. Assuming that all the spins transmitted to Pt are absorbed, the relaxation rate of the YIG becomes

$$\frac{1}{\tau} = \left(\alpha_0 + \frac{\gamma\hbar}{4\pi M_s t_{YIG}} g_{\uparrow\downarrow}\right)\frac{\partial\omega}{\partial H}\frac{\omega}{\gamma}, \qquad (4)$$

where $4\pi M_s$ is the YIG magnetization, $t_{YIG}$ is the YIG thickness, and $\alpha_0$ is the damping parameter of the pure YIG film. It is quite remarkable that the measured values of $g_{\uparrow\downarrow}$ reported in the literature ($\sim 10^{18}$ m$^{-2}$) vary only within an order of magnitude throughout the various surface preparation of the YIG/Pt interface [140]. It is also important to emphasize that this value is about one order of magnitude smaller than the spin mixing conductance reported for all-metallic systems, such as Py/Pt. The other signature that transfer of spin indeed occurs through the YIG/Pt interface is the emission of microwave radiation when the system is pumped out of equilibrium by a dc current [6,7]. Indeed, the spin Hall effect (SHE) [36,37] allows the creation of a pure spin current transversely to the charge current, with an efficiency given by the spin Hall angle $\Theta_{SH}$. Since the spin transfer torque on the magnetization is collinear to the damping torque, there is an instability threshold when the natural damping is fully compensated by the external flow of angular momentum, leading to spin-wave amplification through stimulated emission. The corresponding threshold corresponds to an external flow of spin $J_S^*$ that exactly compensates the flow of spin lost by relaxation:

$$J_S^* = -\left(\frac{1}{\tau} t_{YIG}\right) \frac{M_S}{\gamma} \qquad (5)$$

In their seminal letter to Nature [34], Kajiwara *et al.* showed that by depositing Pt on top of a 1.3 µm thick YIG, one could effectively transfer spin angular momentum through the interface between these two materials. The striking feature was the detection of auto-oscillations of magnetization in YIG by using the inverse spin Hall effect (ISHE) when enough dc current was passed in a Pt electrode. This finding has created a great excitement in the community, although the result turned out to be difficult to reproduce. Threshold effects such as the ones observed in [34] are known to be very sensitive to the exact nature of the magnons being excited [35]. Thus, the central question that remained to be tackled was the degree of coherence of the magnons being excited in the autonomous regime. Of particular concern is the interplay with thermal magnons produced by Joule heating and thermal gradients. In this regard, the characteristic propagation length of the magnons excited by spin transfer should indirectly gives a hint of the energy range of the magnons at play here [141].

To address this fundamental question, a ground effort was started in 2012 to study the effects of spin transfer produced by the spin current on the coherent precession mode of YIG (also known as the Kittel mode). From the beginning it was anticipated that the key to observe auto-oscillations was to minimize the threshold current given by Eq. 5. The first venue is of course to choose a material whose natural damping is very low. In this respect YIG is the optimal choice. The second one is to reduce the thickness of the ferromagnet film $t_{YIG}$, since spin-current induced torque is an interfacial effect. This triggered an effort in the fabrication of ultra-thin films of YIG of very high dynamical quality [132,142,143]. Fig. 7(b) shows a measurement of the ISHE performed on a large slab of 200 nm thick YIG grown by LPE [57]. For such thick YIG layers, the additional broadening produced by the adjacent metallic layer is hardly observable, though, the spin pumping can still be detected through ISHE [144-146]. The inversion of the signal polarity by either inverting the sign of the magnetic field, or by using two metals whose d-shells are either less than half-filled or more than half-filled [57] are all consistent with the claim that the observed electrical signal is produced by the spin Hall effect. The generated voltage being proportional to the length of the sample across the metal, can easily reach several tens of microvolts in millimeter-sized samples. Due to Onsager relations, a charge current injected into Pt should lead to a reduction of the linewidth. But the striking result in Fig. 7(b) was that there were no evidence of auto-oscillations in millimeter-sized samples at the highest dc current possible in the top Pt layer (about $10^{10}$ A/m$^2$) before it evaporates. Even for 20 nm thick YIG films with damping constant as low as $\alpha=2\times10^{-4}$, no measurable effect could be detected on millimeter-size slabs [132]. It is worth mentioning at this point that reducing further the thickness or the damping parameter of such ultra-thin YIG films [147] does not help anymore in decreasing the threshold current, as the relevant value of the damping is that of the YIG/Pt hybrid, which ends up to be completely dominated by the spin-pumping contribution. Figure 7(c) illustrates this statement by calculating the value of the threshold current as a function of the YIG thickness using the parameters listed in Table 1. In this figure, the value of the threshold current only decreases by a mere 10% by decreasing the film thickness from 20 nm to 1.5 nm (one unit cell). This calculation also account for an additional correction factor, $T$ indicating how the transmitted spin current $J_S$ through the interface contributes effectively [148,149] to the charge current $J_C = T J_S \Theta_{SH}$ being converted by the spin Hall angle, $\Theta_{SH}$:

$$T = \frac{\tanh\left(\dfrac{t_{Pt}}{2\lambda_{sd}}\right)}{\coth\left(\dfrac{t_{Pt}}{\lambda_{sd}}\right) + \dfrac{\sigma}{2\lambda_{sd} G_{\uparrow\downarrow}}}, \qquad (6)$$

where $t_{Pt}$ and $\lambda_{sd}$ are respectively the thickness and the spin diffusion length of the Pt layer and $\sigma$ is its electrical conductivity. In the case of magnetic insulators, this correction is significant as it can be as low as $T\approx0.1$ [61]. The reason behind is that, contrary to magnetic metals, where the spins in the accumulation layer relax mostly through spin transfer processes at the NM/FM interface, in the case of YIG the dominant relaxation channels are spin flip events occurring within the Pt spin diffusion length. This is due to the approximately ten times smaller spin mixing conductance at the YIG/Pt interface compared to the Py/Pt one, already pointed out. We have summarized in the table below the typical parameters for the YIG/Pt system.

Table 1.

| Pt | $t_{Pt}$ (nm) | $\rho$ ($\mu\Omega*cm$) | $\lambda_{sd}$ (nm) | $\Theta_{SH}$ |
|---|---|---|---|---|
|  | 7-10 | 17 | 3.5 | 0.06 |
| YIG | $t_{YIG}$ (nm) | $4\pi M_s$ (G) | $\gamma$ (rad s$^{-1}$ G$^{-1}$) | $\alpha_0$ |
|  | 20 | $2.1\times10^3$ | $1.77\times10^7$ | $2-4\times10^{-4}$ |
| YIG/Pt | $T$ | $G_{\uparrow\downarrow}/G_0$ ($10^{18}$ m$^{-2}$) |  |  |
|  | 0.1-0.2 | 4 |  |  |

Overall, the whole process of converting a pure spin current produced in the YIG into a charge current into the Pt (or vice versa) is expected to be about 2 orders of magnitude less efficient than in metals. It turns out that the loss of efficiency roughly offsets the expected gain associated with the decrease of the value of $\alpha_0$ compared to metals. If one would estimate the threshold current required to fully compensate the damping of the FMR mode in typical YIG films [35,47], one would get current densities of about $10^{11}$ A/m$^2$, similar to the values required in fully metallic systems. Thus, the lack of a visible effect on the Kittel mode in Fig. 7(b) is consistent with the estimation of the threshold current for a coherent auto-oscillation.

Concentrating back on the low energy part of the spin-wave spectrum, most notably, none of these high-quality ultra-thin YIG films display a purely homogeneous FMR line. The reason for that is well known. In such extended films, there are many degenerate modes with the main, uniform FMR mode, which through the process of two-magnon scattering broaden the linewidth [150,151]. A striking evidence of these degenerate modes can be obtained by studying more closely the FMR absorption line on the extended film (see in Fig. 8(b)). The lineshape in Fig. 8(b) is neither Lorentzian nor symmetric. This is attributed to inhomogeneous broadening.

Another evidence of the presence of these degenerate modes can be obtained by parametrically pumping the SW modes. This method reveals an uncountable number of modes, which are at the same energy as the FMR mode [144]. Any threshold instability will be affected by the presence of those modes, since mode competition is known to have a strong influence on the emission threshold. Nearly degenerate SW modes compete for feeding from the same dc source of angular momentum, a phenomenon that could become self-limiting and prevent the onset of auto-oscillations. Thus, the next natural step was to reduce the lateral sizes of the YIG film in order to lift the degeneracy between modes through confinement [65]. A series of YIG nanodisks has been subsequently patterned using standard electron lithography and dry ion etching (see Fig. 8(a)). After the YIG lithography, an insulating layer of 50 nm $SiO_2$ was deposited on the whole surface and a 150 nm thick and 5 µm wide microwave Au-antenna was deposited on top. To measure the linewidth of the nanodisks buried under the microwave antenna, a magnetic resonance force microscope (MRFM), which detects the spin-wave absorption spectrum mechanically, was used [152].

The microstructures of YIG revealed that the patterning indeed narrowed the linewidth through a decrease of the inhomogeneous part [153], as shown in Fig. 8(c) and (d). The effect is clear in the perpendicular geometry, where magnon-magnon processes are suppressed owing to the fact that the FMR mode lies at the bottom of the SW dispersion spectrum. However, this is not the case in the parallel geometry, where the FMR mode is not the lowest-energy SW mode.

*4.1. Microwave spectroscopy of auto-oscillations*

The next natural step was to connect these microdisks to two electrodes enabling the injection of a dc current $I_{dc}$ in the Pt layer. The configuration of this experiment is shown schematically in Fig. 9(a) with the presence of a top inductive antenna to generate a uniform microwave field $h_{rf}$. In the following, we report on dynamic studies of the magnetic microdisks with diameter ranging between 2 µm and 5 µm based on a hybrid YIG(20 nm)/Pt(8 nm) bilayer using YIG layer grown by pulsed laser deposition [132].

In order to characterize the flow of angular momentum across the YIG/Pt interface, ISHE-based FMR spectroscopy was performed on the microdisks. The in-plane magnetic field is applied in a transverse direction with respect to $I_{dc}$, as shown in Fig. 9(a). This is the most favorable configuration as far as the symmetry of the SHE spin accumulation is concerned to compensate the damping and to obtain auto-oscillations in YIG, as spins accumulated at the YIG/Pt interface due to SHE in Pt will be collinear with its magnetization. In this configuration, the FMR mode of the YIG disk is excited by the top antenna while the dc voltage across Pt is monitored at zero current (see Fig. 9(a)). As described above, a voltage $V_{ISHE}$ develops across Pt when the FMR conditions are met in YIG. The observed peak corresponds to the excitation of the Kittel mode (uniform precession throughout the disk). In agreement with the symmetry of ISHE, the produced voltage changes sign as the field is reversed, as shown in Fig. 9(b), where the FMR spectra of the 4 µm and 2 µm microdisks are respectively detected at 1 GHz and 4 GHz. From these ISHE measurements, the dispersion relation of the main FMR mode can be determined, as shown in Fig. 9(c). The obtained dispersion relation follows the expected Kittel law.

Spectroscopy of the disks can also be performed independently by MRFM. The FMR spectrum is obtained by recording the vibration amplitude of the cantilever while scanning the external bias magnetic field, $H_0$, at constant microwave excitation frequency. The MRFM setup was used to detect the effect of a dc current, $I_{dc}$, on the FMR spectra when a 5 µm diameter disk is magnetized in-plane by a magnetic field along the +$y$-direction (positive field). The spectra recorded at $f$=6.33 GHz are shown in Fig. 10(a). The middle row shows the absorption at zero current. As the electrical current is varied, a very clear change in the linewidth is observed. At negative current, the linewidth decreases, to reach about half the initial value at $I_{dc}$ = -8 mA. This decrease is strong enough so that the individual modes can be resolved spectroscopically within the main peak. Concomitantly the amplitude of the MRFM signal increases. At positive current, the linewidth increases to reach about twice the initial value at $I_{dc}$ = +8 mA, and the amplitude of the signal decreases. The opposite behavior is observed when the current polarity is reversed (Fig. 10(b)). Hence a clear effect of spin transfer torque acting on the coherent (Kittel) mode is evidenced [61,100]. It is also important to emphasize that for a given field polarity, the product between $V_{ISHE}$ and $I_{dc}$ must be negative to compensate the damping [61], which enables observation of auto-oscillations. This result unambiguously showed that the linewidth in a micron-sized YIG/Pt disk can be tuned by SHE.

Above the threshold dc current, the voltage produced in the inductive antenna by auto-oscillations of the 4 µm and 2 µm YIG disk can be monitored with a spectrum analyzer as a function of the dc current $I_{dc}$ injected into Pt (see Fig. 11). Figures 11(a) and 11(b) respectively present the inductive signal $V_y$ detected in the antenna coupled to these two disks as a function of $I_{dc}$. The configuration is depicted in Fig. 10, with a bias field set to $H$=650 Oe. One can clearly see a peak appearing in the power spectral density close to 3.6 GHz in both cases, at a threshold current of about -13.5 mA in the 4 µm disk and -7.4 mA in the 2 µm disk. These two values correspond to a similar threshold current density in both samples of $(4.4\pm0.2)\times10^{11}$ A/m$^2$. The frequency of these peaks is very close to the expected frequency of the Kittel mode at this bias current (details about the excited mode are given in Section 4.2). This finding is also consistent with the detection of a voltage through an inductive antenna, which is mostly sensitive to the most coherent mode (Kittel mode) and not to the higher-order harmonics. The linewidth of the emission peak, lying in the 10 MHz range (Fig. 11(d)) also proves the coherent nature of the detected signal. As the dc current is varied towards more negative values, the peaks shift towards lower frequency (Fig. 11(c)), at a rate, which is twice faster in the smaller disk. This frequency shift is mainly due to linear and quadratic contributions of Oersted field and Joule heating, respectively. At the same time, the signal first rapidly increases in amplitude, reaches a maximum, and then drops until it cannot be detected anymore, as seen in Fig. 11(e), which plots the integrated power *vs* $I_{dc}$. The maximum of power measured in the 4 µm disk (2.9 fW) is four times larger than the one measured in the 2 µm disk (0.7 fW). This indicates that the same maximum cone angle of precession, θ, is achieved in the two disks. For the same θ, the inductive voltage produced by the 4 µm disk is thus twice larger than by one produced by the 2 µm disk, hence the ratio four in power.

In order to check whether the obtained values of the threshold current $(4.4\pm0.2)\times10^{11}$ A/m$^2$ are consistent with the theoretical expectations, one needs to determine the damping of the disks. From the ISHE measurements, the frequency dependence of the full linewidth at half maximum of the main FMR mode can be determined, as shown in Fig. 11(f). The damping parameters of the 4 µm and 2 µm microdisks extracted from linear fits to the data, $\Delta H=2\alpha\omega/\gamma + \Delta H_0$ (continuous lines in Fig. 11(f), ω is the precession frequency and γ is the gyromagnetic ratio), are found to be similar with an average value of $\alpha=(2.05\pm0.1)\times10^{-3}$. The small inhomogeneous contribution to the linewidth observed in both microdisks, $\Delta H_0$=1.3±0.4 Oe and $\Delta H_0$=0.7±0.4 Oe, respectively, decreases with the diameter and is attributed to the presence of several unresolved modes within the resonance line (see Fig. 10). In order to achieve auto-oscillations, the additional damping term due to the spin-current injection has to compensate the natural relaxation rate $\Gamma_r = 1/\tau$ in YIG following Eq. (5). If only the homogeneous contribution to the linewidth is taken into account, the threshold current $I_{th}$ is thus expected to depend linearly on $H$, as shown by the dashed lines in Fig. 11(g). It qualitatively explains the dependence of $I_{th}$ at large bias fields in both microdisks, but underestimates its value and fails to reproduce the optimum observed at low bias fields. To understand this behavior, the finite inhomogeneous contribution to the linewidth $\Delta H_0$ (Fig. 11(f)) should be considered as well. In fact, this contribution dominates the full linewidth at low bias field. In that case, the expression of the relaxation rate writes:

$$\Gamma_r = \Gamma_G + \gamma \frac{\Delta H_0}{2} \frac{H + 2\pi M_s}{\sqrt{H(H + 4\pi M_s)}}. \qquad (7)$$

Using the value of $\Delta H_0$ extracted from the data of Fig. 11(f), the continuous lines in Fig. 11(g) were calculated, which are in a very good agreement with the experimental data. In this case, both the position of the optimum and the exact value of $I_{th}$ are well reproduced for both disks. Hence, it turns out that quasi-degenerate SW modes, which are responsible for the inhomogeneous contribution to the linewidth, strongly affect the exact value of the threshold current. In fact, it is the total linewidth that truly quantifies the losses of a magnetic device regardless of the nature and number of microscopic mechanisms involved. Even in structures with micrometer-sized lateral dimensions, there still exist a few quasi-degenerate SW modes as evidenced by the finite $\Delta H_0$ observed in Fig. 11(f). Due to the magnon-magnon scattering, these modes are linearly coupled to the main FMR mode, which, as a result, has its effective damping increased, along with the threshold current [76].

The observed behaviors of the integrated power (Fig. 11(e)) and the linewidth (Fig. 11(d)) are probably reminiscent of the presence of these quasi-degenerate SW modes. Moreover, the maximal angle of precession reached by auto-oscillations is found to be about θ=1° in both microdisks [61]. Finally, the disappearance of the signal as $I_{dc}$ gets more negative, is accompanied by a continuous broadening of the linewidth, which increases from a few MHz to several tens of MHz (Fig. 11(d)). All these behaviors are consistent with a saturation of the resonance. It is known that such phenomena occur in the high-power dynamics of the FMR excitation (also called Suhl 2$^{nd}$ order instability) [113]. When the precession angle exceeds a certain value, the Kittel mode decays very efficiently into degenerate modes. The value of the coupling to the external source is such as to keep the cone angle constant at a threshold amplitude through an increase of the broadening. This process is very efficient for very low damping materials, and it occurs when the number of coherent magnons created is the same as the one that would be excited by an rf field of the order of the linewidth. A meaningful interpretation of these experimental results is that as the FMR mode starts to auto-oscillate and to grow in amplitude as the dc current is increased above the threshold, its coupling to other SW modes - whose amplitudes also grow due to the spin-current induced torque - becomes larger, which makes the flow of energy out of the FMR mode more efficient. This reduces the inductive signal, as non-uniform SW modes are poorly coupled to the inductive detection scheme. At the same time, it enhances the auto-oscillation linewidth, which reflects this additional nonlinear relaxation channel.

*4.2. Dynamic modes excited in magnetic insulators by pure spin current*

Although the microwave-spectroscopy measurements have clearly demonstrated auto-oscillations in YIG-based devices [76], they could not provide sufficient information about the dynamic modes excited in the system by the pure spin current. In particular, based on the results of the electronic measurements, it is difficult to conclude whether the auto-oscillation mode is a self-localized spin-wave bullet, which was observed in qualitatively similar all-metallic nano-gap devices (Section 3.1), or it is some other, qualitatively different auto-oscillation mode created due to the specific nonlinear response of low-damping YIG. This issue was addressed by the direct spatially-resolved imaging of the dynamic modes utilizing micro-focus BLS technique adapted for measurements with ultra-thin YIG films [77].

Figure 12(a) shows the schematic of the BLS experiment utilizing samples described in detail in the previous Section. Since the optical access to the YIG disk in the used devices is blocked by the Pt layer and Au electrodes, the probing laser light was sent through the GGG substrate and was focused by a specially corrected microscope objective lens into a diffraction-limited spot on the YIG/Pt surface. The wavelength of the laser was chosen to be 473 nm, which provided high sensitivity of the method for measurements with ultra-thin YIG. The power of the probing light was as low as 0.05 mW, which guaranteed negligible laser-induced heating of the sample. The modified micro-focus BLS technique provided the opportunity to record spectra of magnetic fluctuations in the YIG disk and analyze their modification under the influence of the pure spin current in the same way, as it was done for all-metallic SHE systems. These measurements revealed an enhancement of magnetic fluctuations in the YIG disk at low currents followed by an onset of the current-induced auto-oscillations at $I = I_C \approx 8$ mA. At this current, a narrow peak with the intensity exceeding that of the thermal fluctuations at $I = 0$ by about two orders of magnitude was found to emerge in the BLS spectrum [77].

The information about the dynamical modes contributing to the current-induced auto-oscillations was obtained by two-dimensional mapping of the dynamic magnetization in the YIG disk, as illustrated in Fig. 12(b). As seen from these data, the auto-oscillations are strongly localized in the $y$-direction in the area corresponding to the gap between the electrodes, whose edges are marked in Fig. 12(b) by the horizontal lines. In contrast, there is no pronounced localization of the auto-oscillations in the $x$-direction. They clearly extend over the entire active device area, where the injection of the spin current takes place. Figure 12(c) illustrates the modifications of the spatial profiles of the auto-oscillations with current. While the $y$-profiles are nearly independent of $I$, $x$-profiles exhibit a strong current dependence. In particular, one clearly sees that the auto-oscillations do not always occupy the entire active device area. At currents close to $I_C$, they are strongly localized in the middle of the gap. However, with the increase in $I$, they rapidly expand in the $x$ direction. This process is further characterized in Fig. 12(d), which shows the current dependence of the full width at half maximum of the $x$-profiles. This width linearly increases from 0.5 μm to about 1.4 μm for $I$ varying from 8 to 9 mA and then saturates at the value of 1.5 μm.

We emphasize that this variation of the size of the auto-oscillation mode is in contradiction with the typical manifestations of the dynamic magnetic nonlinearity, which is expected to result in the spatial self-localization of the intense oscillations and lead to a formation of the spin-wave bullet, as was observed in all-metallic SHE systems (Section 3.1). Note that the spatial dimensions of the bullet mode are not determined by the dimensions of the area where the spin current is injected, but are governed by the self localization phenomena [118]. In contrast, in the studied system, the auto-oscillation mode tends to extend over the entire active area indicating that the self-localization does not play an essential role. The absence of the self-localization in YIG can be attributed to a nonlinear limitation of the amplitude of the auto-oscillations. Indeed, if this amplitude is limited by any other nonlinear phenomenon, the self-localization mechanism requiring the dynamic magnetization to be comparable with the static magnetization can be suppressed.

This scenario is supported by the data of Fig. 12(e). The maximum intensity of the auto-oscillations (point-down triangles) saturates immediately after the onset and then stays nearly constant over the entire used range of $I$. In contrast, the integral intensity (point-up triangles) exhibits a gradual increase up to $I$ = 9.2 mA. This indicates that the amplitude of the dynamic mode excited at the onset cannot grow above a certain level and that the further increase in $I$ results in the energy flow into other dynamical modes leading to the formation of a strongly nonlinear spatially-extended mode. Note that similar effects closely related to a small magnetic damping in YIG have been already observed in [154]. The small damping leads to a low onset threshold for nonlinear spin-wave instabilities [113,155] resulting in the efficient energy transfer from the auto-oscillation mode to the short-wavelength spin-wave modes. These processes cause the saturation of the intensity of the auto-oscillation mode. Moreover, they result in the reduction of the absorption by the auto-oscillating mode of the angular momentum brought into the system by the spin current [156]. As a result, the injected angular momentum gets channeled into long-wavelength modes at frequencies close to that of the fundamental mode excited at the auto-oscillation onset, which manifests itself as a spatial and spectral broadening of the auto-oscillations.

The spatially-resolved BLS measurements also shed light on the nature of the mode, which is excited in the YIG disk at $I = I_\text{C}$ before the nonlinear broadening takes place. Since the nonlinear self-localization effects do not play a significant role in the studied system, one expects that the spin current will excite the fundamental quasi-uniform ferromagnetic-resonance (FMR) mode of YIG disk. However, as seen from the data of Fig. 12(c), at $I = I_\text{C}$, the auto-oscillation mode shows a strong localization at the center of the disk, which does not allow one to identify it as the FMR mode. This inconsistency was resolved by BLS measurements at $I < I_\text{C}$, in which the FMR in the YIG disk was excited by a dynamic magnetic field created by an additional stripe antenna aligned parallel to the gap between the electrodes. Figure 12(f) shows the spatial maps of the field-driven FMR mode recorded by BLS at $I = 0$ and 7 mA. The maps clearly demonstrate that the FMR mode initially spread over the entire disk becomes localized at its center when the dc current is applied. Since the measurements were performed at sufficiently small microwave powers to exclude any nonlinear effects [157], the observed localization can only be attributed to the joint effect of the Oersted field of the dc current, the local reduction of the magnetization due to the Joule heating, and the partial compensation of the dynamic damping by the pure spin current injected into the YIG film. This assumption was proven by micromagnetic simulations which reproduced well the experimental observations [77]. These results clearly showed that the mode excited in the YIG disk at the auto-oscillation onset is the confined FMR mode modified due to the side effects of the electrical current flowing in the Pt layer.

We would like to emphasize that the results described above are the first results on the excitation of coherent magnetization dynamics in magnetic insulators by pure spin current. Undoubtedly, further studies are needed before one can make a conclusion about the potential of small-damping magnetic insulators for spin-torque devices. However, it is clear already now, that, contrary to initial optimistic expectations, the small damping of YIG can also play a negative role by stimulating undesirable nonlinear processes resulting in the saturation effects and the lost of coherence of the current-induced auto-oscillations. It is, therefore, decisive for the future developments to find out how these processes can be suppressed, for example, by controlling the spectrum of the spin-wave modes in nano-scale devices.

## 5. Excitation of magnetization oscillations and waves by nonlocal spin injection

The practicality and efficiency of spin Hall mechanism for the excitation and control of magnetization dynamics in conducting and insulating magnetic materials has been already convincingly demonstrated. However, studies have also revealed several shortcomings of this mechanism. The most significant issue is the strong Joule heating due to the flow of the driving electric current in high-resistivity SHE materials. This is particularly detrimental for the YIG-based devices [76,77,83], since the magnetization of YIG is strongly dependent on temperature because of its low Curie point [129]. In particular, the heating processes are intrinsically slow, resulting in a long settling time of the stationary oscillation regime. Thus, SHE-based oscillators are not well-suited for the generation of short dynamic magnetization pulses in applications associated with high-speed data transmission and processing. Finally, the increase of temperature in the auto-oscillation area unavoidably leads to a decrease of the magnetic auto-oscillation coherence, due to the increased thermal noise.

In this Section, we discuss an alternative mechanism for the generation of pure spin currents – nonlocal spin injection (NLSI) – which allows one to overcome the difficulties described above. This mechanism relies on the spatial separation of charge and spin currents by providing an additional current path that bypasses the active magnetic layer [40,41]. NLSI has been traditionally studied in lateral spin-valve structures, where the spin injector and the spin absorber (the active magnetic element) are laterally separated by the distance of several hundred nanometers, which compromises the device efficiency due to the large spin-diffusion losses. More recently, it was realized that the efficiency of spin-current generation by NLSI can be significantly increased by utilizing vertical spin valves with a modified structure [85,86]. In the vertical NLSI devices, a relatively thick normal-metal layer carrying the electrical current is simply sandwiched between two thin ferromagnetic films. One of them plays the role of the spin injector, and the other is the active magnetic layer. It was recently shown that NLSI devices based on this principle are capable of highly coherent magnetization auto-oscillation characterized by the spectral linewidth of a few megahertz at room temperature [87,88]. The simple and flexible layout of the NLSI devices enables a straightforward implementation of mutually synchronized NLSI oscillators [89], as well as devices capable of directional emission of coherent propagating spin waves [23], which has not yet been achieved in the SHE-based devices because of the limitations imposed by their geometry. Moreover, the NLSI devices have been found to exhibit negligible Joule heating [23] and a very fast response to the driving current, enabling generation of short pulses of the dynamic magnetization and short spin-wave packets [90].

*5.1. Nonlocal spin-injection nano-oscillators*

The schematic of the NLSI oscillator is shown in Fig. 13(a). It is formed by a 60 nm circular nanocontact on an extended multilayer that consists of a 5 nm thick Permalloy (Py) active magnetic film separated from the 8 nm thick CoFe spin injector by a 20 nm thick layer of Cu. The driving electric current $I$ is injected into the multilayer through the nanocontact and is then drained to the side electrodes (not shown). Because of the large difference in the resistivities of the layers comprising the device, most of the current is drained through the Cu film, while only about 3% of the current is shunted through the active Py layer [23]. The red arrows in Fig. 13(a) show the corresponding flow of electrons. The injected electrons become spin polarized due to the spin-dependent scattering in CoFe and at the Cu/CoFe interface [158], resulting in spin accumulation in Cu above the nanocontact. Spin diffusion away from this region produces a spin current that flows into the Py layer and exerts STT on its magnetization. The magnetizations of both the CoFe and the Py layers are aligned with the saturating static in-plane magnetic field $H_0$. For positive driving electric currents, as defined in Fig. 13(a), the magnetic moment carried by the spin current is antiparallel to the magnetization of the Py layer, resulting in the STT compensating the dynamic magnetic damping. When damping becomes completely compensated by the spin current, the magnetization of the Py layer starts to auto-oscillate in the area above the nanocontact [87].

Figure 13(b) illustrates the spin current-induced enhancement of magnetic fluctuations and the onset of magnetization auto-oscillations in the NLSI device. The observed behaviors are very similar to those typical for the nanoconstriction SHE devices (compare with Fig. 4(d)). When a small current is applied, low-frequency fluctuations become preferentially enhanced, and a spectral peak gradually emerges in the low-frequency part of the fluctuation spectrum. The intensity of the peak abruptly increases above the critical current $I_C \approx 4.7$ mA, marking the onset of auto-oscillations. The enhancement of the magnetization fluctuations and the onset of auto-oscillations were observed only at $I>0$, while the fluctuations were found to be increasingly suppressed with increasing magnitude of $I<0$. These behaviors were independent of the direction of the static magnetic field $H_0$, consistent with the expected symmetry of the NLSI mechanism and of the effects of STT. The auto-oscillations were observed over a wide range of the static field $H_0$ (Fig. 13(c)), indicating that the mechanism of auto-oscillation in the NLSI oscillators is highly robust. As the field is increased from 500 Oe to 2 kOe, the oscillation frequency (point-up triangles in Fig. 13(c)) monotonically increases by more than a factor of two from 5.8 GHz to 12.4 GHz. The variation of the static field has a minimal effect on the onset current (diamonds in Fig. 13(c)), which is promising for the applications requiring wide-range frequency tunability. The auto-oscillation frequency is lower than the FMR frequency (point-down triangles) at any field, indicating that the auto-oscillation mode is localized. However, the magnitude of the frequency shift is much smaller than in the nano-gap SHE oscillators (Fig. 3(e)). We note that the latter was controlled by the nonlinear properties of the self-localized bullet mode.

In addition to the small frequency shift, several other features distinguish the oscillation mode in the NLSI oscillators from the bullet mode. In particular, the spectral peak corresponding to the auto-oscillation mode does not abruptly appear at the onset of auto-oscillations, but rather gradually emerges at $I<I_C$ (Fig. 13(b)), similarly to the nanoconstriction SHE oscillators (Fig. 4(d)). In contrast, the self-localized bullet mode in nano-gap SHE oscillators is spontaneously formed at the onset of auto-oscillations. The difference with the bullet mode is also highlighted by the two-dimensional BLS mapping of the auto-oscillations (Fig. 13(d)), which yields a half-maximum width of the auto-oscillation mode of about 300 nm, significantly larger than expected for self-localized oscillations. Furthermore, at large currents the NLSI oscillators transit to the multimode oscillation regime characterized by close frequencies of the auto-oscillation modes, which is not expected for the bullet-mode dynamics. All of these observations suggest that the auto-oscillation modes of NLSI oscillators are localized by an effective potential well, similarly to the nanoconstriction SHE oscillators. In the SHE devices, the potential well was formed due to the dipolar field of the patterned magnetic film. However, this mechanism cannot be at play in the NLSI devices based on extended magnetic films. Moreover, the symmetry of the NLSI devices ensures that the Oersted field of the current is negligible in the active magnetic layer, and therefore it also cannot be responsible for the localization.

It was shown in Ref. [87] that the experimentally observed characteristics of the NLSI oscillators can be accurately described by a quasi-linear model that takes into account the enhancement by the spin current of the magnetic fluctuations in the Py film above the nanocontact. As was discussed in Section 2.2, the spin current-induced fluctuation enhancement can result in the reduction of the effective magnetization $M_e$ by more than 50% (Fig. 2(d)). In the NLSI devices, the spin current is locally injected into the Py film, creating an inhomogeneous distribution of $M_e$ that can confine spin-wave modes. Curves in Fig. 13(e) show the results of the micromagnetic simulations demonstrating two confined modes in an effective potential well formed due to this mechanism, and symbols in this Figure show the experimentally determined frequencies of the auto-oscillation modes. The calculation is based on a linear approximation for the dependence of the effective magnetization on the applied current, $M_e=M_0-aI$. A single fitting parameter $a$ describing the dependence of magnetization on current was sufficient to achieve an excellent agreement with the experiment for the frequencies of both modes and for their dependencies on current, confirming the validity of the model.

The spatial characteristics of the auto-oscillation mode are controlled by the geometry of spin current injection into the Py layer, which is in turn determined by the spin diffusion properties of the nonmagnetic spacer separating this layer from the CoFe spin injector. A large spin diffusion length in Cu leads to a large size of the auto-oscillation mode (see Fig. 13(d)), minimizing the effects of thermal noise on the coherence of auto-oscillations in the NLSI oscillators. Since the BLS measurements cannot provide accurate information about the auto-oscillation spectral linewidth, the latter was analyzed by the electronic microwave spectroscopy [88]. In these experiments, an additional inductive antenna was integrated into the NLSI devices. The oscillating dipolar field produced by the magnetization auto-oscillations induced a microwave current in the antenna. The induced microwave signal was amplified and detected by a spectrum analyzer. Figure 13(f) shows a representative spectrum of the electronic signal generated by the oscillator. These data demonstrate that the quality factor of oscillations in the NLSI devices is significantly higher than in the SHE oscillators (compare with Fig. 4(b)). This is not surprising, since in addition to the reduced thermal noise effects due to the larger oscillation area, the operation of NLSI oscillators does not require the flow of the electrical current through highly-resistive materials like Pt, resulting in a significantly smaller Joule heating. This further reduces the detrimental effects of thermal noise on the oscillation coherence.

The thermal advantages of NLSI-based devices are highlighted by Fig. 13(g), which shows the current dependence of the temperature increase in the active area of the NLSI (point-up triangles) and in the nanoconstriction SHE (point-down triangles) oscillators, calculated based on the heat flow simulations using COMSOL Multiphysics v5.0 software (COMSOL Inc.) (see [23] for details). As seen from these data, even at $I=I_C$, at the onset of auto-oscillations, the temperature increase in the active area of the SHE devices exceeds 10 K, while for the NLSI oscillators this increase does not exceed 1.5 K. The difference becomes even more dramatic at larger currents. For instance, at $I=2I_C$ the temperature in the SHE devices increases by more than 50 K, while the corresponding increase in the NLSI oscillators is less than 6 K, which is not expected to significantly affect the oscillation characteristics even in YIG-based devices. Such a small temperature increase is achieved in the NLSI oscillators not only because of lower heat generation in the current-carrying layers, but also due to the more efficient heat conduction away from the active device area by the thick Cu spacer.

*5.2. Mutual synchronization of NLSI oscillators*

The extremely simple layout of the NLSI oscillators and the large spatial dimensions of their auto-oscillation area facilitate the development of complex magnetic nano-circuits based on these devices. For example, a simple modification of the layout shown in Fig. 13(a) to include several nanocontacts separated by the distance of several hundred nanometers can be used to form a system of interacting NLSI oscillators. Thanks to the large size of the auto-oscillation area, the interaction between oscillators in such a system can be mediated by the direct spatial overlap of the auto-oscillating modes of the neighboring devices, resulting in their efficient dynamic coupling. Such ensembles of interacting oscillators are promising for advanced applications such as magnetic neural networks and logic circuits [14]. A more immediate application for the dynamic interaction is mutual synchronization of multiple oscillators (see also Section 3.3). Note that, in contrast to the SHE-based devices, the implementation of such systems with NLSI oscillators does not require patterning of the active magnetic layer, which allows one to arrange coupled oscillators in two-dimensional arrays. Moreover, in contrast to the traditional STT oscillators operating with spin-polarized electric current, the active magnetic layer remains fully accessible to spatially-resolved magneto-optical measurements, and its topography can be easily modified to achieve the desired dynamic and spatial characteristics of the oscillators.

Mutual synchronization of NLSI oscillators was demonstrated in [89]. The system studied in this work is similar to that shown in Fig. 13(a), but instead of a single nanocontact it comprises two 60 nm nanocontacts fabricated at a distance of 400 nm from each other. The driving electric current $I$ is injected into the multilayer through both nanocontacts simultaneously, forming a system of two NLSI oscillators connected in parallel. Figure 14(a) shows the spatial map of the spin current-induced magnetization dynamics in such a system recorded by BLS at $I$=20 mA. The map clearly shows two spatially overlapping oscillating magnetization regions centered on the point contacts labeled "A" and "B". At small currents, only one of the devices exhibits oscillations, while the second remains in the sub-critical regime, likely due to the slight difference in their size. From measurements in this regime, one can estimate that the full width of the auto-oscillation area at half-maximum intensity is 350 nm in the direction along the line connecting the two nanocontacts. Since this dimension is close to the separation between the contacts, the oscillating magnetization regions should exhibit a significant spatial overlap, resulting in efficient interaction between the oscillators.

The interaction was analyzed based on independent measurements of the oscillation characteristics of the two devices. This was accomplished by placing the probing laser spot at the location of the corresponding nanocontact, and recording the BLS spectrum at this position. The dependencies of the frequencies of the two oscillators on dc current were quantified by these measurements (Fig. 14(b)). At small currents, the auto-oscillation frequencies differ by 0.3-0.4 GHz, which is likely caused by the slightly different diameters of the nanocontacts. At a current $I_{SYN}$=18 mA, the frequencies of both oscillators abruptly jump and become equal, indicating mutual synchronization of the two devices. As the current is further increased, both frequencies exhibit a nonlinear redshift but remain equal. Finally, starting from the current $I_L$=23 mA, they gradually diverge, indicating the loss of synchronization. In contrast to the synchronization onset, the loss of synchronization at large current is not abrupt, but is characterized by a gradual increase of the difference between the frequencies of the two oscillators. This indicates that the synchronization is not completely lost at large currents, but is rather increasingly disturbed with increasing $I$.

The dependence of the oscillation characteristics on the applied field provides further insight into the mechanisms of synchronization. The field dependences of $I_{SYN}$, $I_L$, and of the auto-oscillation onset current $I_C$ of oscillator A are summarized in Fig. 14(c). The synchronization onset current monotonically decreases with increasing field, while the synchronization loss current increases, resulting in an overall significant increase of the synchronization current interval. In contrast, the auto-oscillation onset current exhibits a weak non-monotonic dependence on field. Thus, one can conclude that the variations of $I_{SYN}$ and $I_L$ are not associated with a simple rescaling of the characteristic oscillation currents. The observed dependencies clearly indicate that the synchronization is facilitated by the field increase.

Analysis performed in [89] showed that the observed behaviors can be explained by the competition between the dynamic nonlinearity that facilitates synchronization, and the spin current-enhanced short-wavelength magnetic fluctuations that suppress synchronization. In particular, the onset of synchronization was found to be closely correlated with the transition of the oscillators into the strongly nonlinear oscillation regime. This correlation is in agreement with the established theories of synchronization of STT devices [92]. The gradual loss of synchronization at large current was explained by the increasing enhancement of magnetization fluctuations by the spin current. Because of the nonlinear coupling among different dynamical modes, enhanced fluctuations affect the phase stability of the oscillators causing a partial loss of their synchronization. The fluctuations are not expected to significantly depend on the field, because they are dominated by the large phase volume of short-wavelength modes whose spectral characteristics are mostly determined by the exchange interaction, and not by the modest static field. In contrast, the nonlinearity of the dynamical states of in-plane magnetized films significantly increases with increasing static field [98]. Since the nonlinearity facilitates synchronization, this results in an increase of the synchronization current interval with increasing field.

The results described above clearly show that the NLSI oscillators are capable of efficient mutual synchronization and that the synchronization can be controlled by the variation of the experimental parameters. These results provide a foundation for future studies exploring new directions for further improvement of the dynamical characteristics of interacting spin-current oscillators, which will likely lead to novel advanced applications.

*5.3. Excitation of propagating spin waves by pure spin currents*

Excitation of spin waves by pure spin currents can have a particularly significant impact on the development of magnonics [20,24-30], which utilizes propagating spin waves as the nanoscale signal carrier. Control of the magnetization dynamics by pure spin currents is especially attractive for nanoscale magnonic devices, since the traditional inductive excitation method becomes very inefficient on the nanoscale [159,160]. In spite of this, excitation of coherent propagating spin waves by pure spin currents was achieved only recently. The difficulty was associated with a number of conflicting requirements, which are also known for the traditional devices operated by spin-polarized electric currents [16-22]. In particular, efficient excitation of single-frequency coherent magnetization oscillations generally requires that they are spatially confined, as was shown for all the demonstrated SHE- and NLSI-based nano-oscillators. Equivalently, the dynamical mode that enters the auto-oscillation regime under the influence of spin current does not radiate energy in the form of propagating spin waves. Moreover, since the spin torque effect underlying the spin current-induced dynamics is exerted only at the magnetic interfaces, maximizing the effect of spin current requires that the thickness of the active magnetic layer does not exceed a few nanometers. However, spin waves rapidly decay in thin magnetic films. Thus, relatively thick active magnetic layers must be utilized to achieve propagation length of several micrometers acceptable for integrated magnonic circuits [28]. It was recently shown [23] that the NLSI spin-current generation mechanism provides the geometric flexibility required for the implementation of a spin current-driven nanomagnonic system that simultaneously satisfies the conflicting requirements described above. This was accomplished by hybridizing two magnetic subsystems with different dynamic characteristics: an active subsystem in which a spatially confined dynamical mode is excited by the spin current, and a spin-wave guiding subsystem that facilitates efficient propagation of spin waves.

The schematic of such a system is shown in Fig. 15(a). Its active part consists of an NLSI nano-oscillator based on the Py(5 nm)/Cu(20 nm)/CoFe(8 nm) multilayer with the 60 nm circular nanocontact fabricated on the CoFe side. Similarly to the devices described in Sections 5.1 and 5.2, the oscillator exhibits spatially localized spin current-induced auto-oscillations in the Py layer above the nanocontact. The typical size of the auto-oscillation area of 300-400 nm is determined by the spin current injection region. To convert these localized magnetization oscillations into a propagating spin wave with a sufficiently large propagation length, a 20 nm thick and 500 nm wide Py strip aligned perpendicular to the direction of the static field $H_0$ is fabricated on the surface of the extended Py film. The waveguide is terminated at a distance of 150 nm from the centre of the nanocontact. This distance is sufficiently small to ensure efficient dynamic coupling between the current-induced magnetic auto-oscillations in the thin film and the magnetization in the strip. Because of this coupling, the localized oscillations are expected to excite a propagating spin wave, provided that there are available spin-wave spectral states at the frequency of the auto-oscillations.

Figure 15(b) shows the dispersion spectra of spin waves in an extended Py film with the thickness of 5 and 25 nm (solid curves) and for spin waves propagating in the stripe waveguide (symbols) calculated for $H_0$=1000 Oe. As seen from these data, the slope of the dispersion curve, which is proportional to the spin-wave group velocity, is significantly larger for thicker extended film. Since the group velocity determines the propagation length, the latter increases from less than 1 μm for a 5 nm film to several micrometers for a 25 nm thick film. The slope of the dispersion curve calculated for the 20 nm thick stripe waveguide manufactured on top of 5 nm thick extended Py film is close to that of the thick extended Py film. Therefore, one can also expect a large propagation length for the spin waves in the waveguide. A distinct feature of the dispersion spectrum of spin waves in the waveguide, which is of particular importance for the possibility to achieve spin-wave radiation, is its significant frequency downshift in comparison with the curves for an extended film. This downshift originates from the reduction of the internal static magnetic field in the waveguide caused by the dipolar field $H_D$ produced by the uncompensated magnetic charges at the waveguide edges (see inset in Fig. 15(a) and Ref. [161]). The downshift of the spin-wave spectrum results in the appearance of spin-wave spectral states at frequencies below the FMR frequency (dashed line in Fig. 15(b)), which enables the radiation of spin waves by the localized current-induced auto-oscillations, even though the frequency of the latter is always smaller than the FMR frequency (Section 5.1).

The auto-oscillation characteristics of the NLSI oscillator integrated into the hybrid spin-wave device described above are summarized in Fig. 15(c), which shows the current dependence of the frequency and the intensity of the auto-oscillations obtained from BLS measurements with the probing laser spot located directly at the position of the nanocontact. The integrated oscillator transits to the auto-oscillation regime at the critical current $I_C$=3.6 mA, which is close to that for stand-alone devices (Section 5.1). This indicates that the oscillation characteristics are not adversely affected by the integration of the oscillator into the magnonic system. The intensity of the auto-oscillations gradually increases with increasing $I>I_C$, while their frequency decreases due to the nonlinear frequency shift. We emphasize that the auto-oscillation frequency is below the FMR frequency in the extended Py film (horizontal dashed line in Fig. 15(c)) within the entire used range of current. As follows from the data of Fig. 15(b), there are no propagating spin-wave states available in the Py film at the frequency of auto-oscillation, and therefore the oscillation does not radiate spin waves into the surrounding Py(5 nm) film. In contrast, the auto-oscillation frequency range is well matched with that of the waveguide mode, which should result in an efficient radiation of spin waves into the waveguide.

Figure 15(d) shows the spatial BLS intensity map recorded by rastering the probing laser spot over a 3 μm by 1.2 μm region encompassing the nano-oscillator and the adjacent area of the waveguide. It clearly shows two merged but distinct dynamical regions. The first circular high-intensity region is centered on the nanocontact. In this region, the magnetization oscillations are excited by the spin current. It is merged with another arrow-shaped increased-intensity region aligned with the strip waveguide, which is indicated by a dashed contour in Fig. 15(d). The increased intensity is entirely confined to the waveguide, as shown by the transverse section of the map, inset in Fig. 15(e). These observations are consistent with the directional propagation of a spin wave excited in the waveguide by the spin current-induced oscillations.

The propagation characteristics of the excited spin wave can be analyzed based on the measured dependence of the BLS signals on the propagation coordinate $x$, which is defined as the position along the waveguide strip with the origin at the location of the nanocontact. The point-down triangles in Fig. 15(e) show the BLS intensity integrated across the transverse sections of the intensity map. These data plotted on the logarithmic vertical scale show that the excited spin wave exhibits a well-defined exponential decay $\sim\exp(-2x/\lambda)$ along the waveguide. Here, $\lambda$ is the propagation length – the distance over which the spin wave amplitude decreases by a factor of e in the propagation direction. By fitting these data with the exponential function (red curve in Fig. 15(e)), one obtains $\lambda=3.0$ μm, which is sufficiently large for the practical implementations of magnonic nano-systems.

The BLS data also allow one to determine the efficiency of spin wave excitation in the waveguide due to the dynamical coupling to the nano-oscillator. One can extrapolate the exponential spin-wave decay curve to the position $x=150$ nm corresponding to the edge of the waveguide, and find the ratio between this value and the intensity at the position $x=0$ of the nanocontact, which characterizes the energy of the localized auto-oscillation mode. From the data of Fig. 15(e), one obtains the coupling of about 35%, demonstrating the high efficiency of the proposed system.

Figure 15(f) shows the dependence of the spin wave propagation length and the coupling efficiency on the driving current. Both of these parameters slightly decrease with increasing *I*. Nevertheless, the decrease of both parameters is moderate, allowing one to tune the operating frequency of the device not only by the static field, but also by varying the driving current. The observed decrease in the propagation length is associated with the nonlinear redshift of the auto-oscillation frequency. As seen from Figs. 15(b) and 15(c), at large currents the auto-oscillation frequency approaches the bottom of the spin-wave dispersion at 7.5 GHz, where the slope of the dispersion curve noticeably reduces, resulting in the reduction of the spin-wave group velocity. We note that the discussed devices can be easily optimized to enhance the propagation length and minimize its dependence on current. In particular, by varying the geometrical and material parameters of the waveguide, one can achieve a stronger shift of the waveguide mode to lower frequencies. In this case, the bottom of the spin-wave dispersion in the waveguide will be located far below the operational frequency range of the oscillator, and the flattening of the dispersion near the bottom will not influence the propagation characteristics of the radiated spin waves. The layout of the NLSI devices also allows one to control the auto-oscillation frequency of the oscillator by using additional magnetic patterns that modify the static magnetic field at the location of the nanocontact [23]. Together with the ability to vary the waveguide parameters, this enables a complete control of the frequency matching between the oscillator and the radiated spin waves.

Finally, we discuss the power consumption and the operational speed of the proposed devices. As was found in [23], the total electrical power consumed by the devices does not exceed 1-2 mW within the entire used range of driving currents. Since the devices perform all the functions necessary for the magnonic operation, ranging from the conversion of the dc current into microwave oscillations to generation of propagating spin waves, one can conclude that their power efficiency is superior to the magnonic systems that utilize spin-wave excitation by the microwave currents generated by the traditional external microwave sources. Moreover, the devices have been found [90] to exhibit very fast response to the driving current pulses, which allows one to excite spin-wave packets with duration down to a few nanoseconds. This eliminates the need for the relatively slow and power-consuming semiconductor microwave switches for the implementation of high-rate magnonic systems for data transmission and processing.

In addition to magnonic nano-circuits, propagating spin wave generation by the NLSI-based devices can also be utilized for the dynamical mutual coupling in systems of nano-oscillators. In contrast to coupling mediated by direct overlap of the auto-oscillating modes (Section 5.2), interaction via propagating spin waves opens extensive possibilities for the control of coupling by varying the wavelength and the phase of spin waves. Together with the possibility to excite short spin-wave packets, this allows the implementation of fast nano-magnonic systems with novel complex functionalities.

## 6. Outlook: perspectives and applications

In this review, we have described the recent advancements in the excitation of magnetization dynamics by pure spin currents. While this research field has emerged only recently, the results discussed in this review illustrate the fast pace of its evolution in the last decade, and the rapidly increasing viability of technical applications. The possibility to excite coherent magnetization oscillations and waves by pure spin currents has been by now firmly established in a number of experiments based on several different approaches. The results have revealed the fundamental mechanisms enabling these processes, although the understanding of physics underlying the observed phenomena remains largely qualitative. In particular, the experiments clearly demonstrated the important role of dynamic nonlinear interactions in the spin system driven into a strongly non-equilibrium state by the spin current. Simple qualitative understanding of this fact enabled the development of spin-current systems exhibiting single-frequency auto-oscillations, and a significant improvement of their oscillation characteristics. However, the details of the nonlinear processes involved in spin current-induced dynamics are still unknown. This is not surprising, since nonlinear spin-wave interactions is an extremely challenging problem even for homogeneous bulk magnetic materials [155]. The development of a rigorous theory capable of describing nonlinear spin-wave interactions in confined systems subjected to spin current will likely significantly advance the research on spin current-induced phenomena, open new directions and lead to novel applications.

The described effects of pure spin currents elucidate a strong connection between the research field of spin current-induced magnetization dynamics and many other research fields in dynamic magnetism. For example, it was recently suggested that the emergence of coherent auto-oscillations in systems driven by pure spin currents can be treated as Bose-Einstein condensation of magnons [162,163]. This emerging understanding establishes a link to the fascinating field of room-temperature magnon condensation, which has been intensely explored over the last decade [106,112,164]. Another closely related research area that gained a significant attention in the recent years is spin caloritronics [165]. The importance of the short-wavelength (thermal) magnon dynamics is already well recognized in this research field. In this review, we have described the spin current-induced excitation of short-wavelength modes as mostly a parasitic effect that suppresses the excitation of coherent auto-oscillations. On the other hand, interaction of spin current with magnetization is one of very few mechanisms that allow one to controllably excite such magnons and explore their propagation characteristics [141,166]. We also emphasize the particular importance of the advancements in spin current-induced excitation of propagating spin waves for the development of the research field of nano-magnonics, which until recently has evolved independently from the field of spin-torque phenomena. While the traditional spin-torque devices were found to be capable of spin-wave excitation [20], the limitations imposed by their geometry have discouraged researchers from using them in magnonic circuits. We believe that the integration of spin-torque and magnonics devices will dramatically accelerate following the recent demonstration of spin-wave excitation by pure spin currents, in a flexible and simple spin-current device structure [23,90]. An additional encouraging benefit of pure spin currents for magnonics is the possibility to excite spin waves in magnetic insulators such as Yttrium Iron Garnet, which is presently viewed as the most suitable material for future nano-magnonic circuits.

To conclude, the recent advancements in the studies of the effects of pure spin currents on the dynamic magnetization in conducting and insulating magnetic materials create new opportunities for many research fields in magnetism, and provide a viable solution for the real-world technical applications of magnetic nano-devices in communication, data storage, and computation technologies.

We would like to acknowledge D. Baither, N. Beaulieu, J. Ben Youssef, R. Bernard, P. Bortolotti, R. Cao, M. Collet, O. d'Allivy Kelly, B. Divinskiy, E. R. J. Edwards, M. Evelt, K. Garcia-Hernandez, S. V. Gurevich, C. Hahn, A. Hamadeh, R. Liu, R. D. McMichael, H. Meley, A. H. Molpeceres, M. Muñoz, V. Naletov, J. L. Prieto, B. Rinkevich, A. V. Sadovnikov, G. Schmitz, A. Slavin, M. D. Stiles, A. Telegin, V. Tiberkevich, H. Ulrichs, M. Viret, and A. Zholud for their contributions to this work.

**Figure captions**

Fig. 1 Schematic of a simple SHE device. Inset illustrates the generation of pure spin current due to SHE in the Pt microstrip. (b) Representative FMR curves recorded by BLS at different currents in the Pt microstrip, as labeled. (c) FMR frequency vs current, measured in the continuous- (point-down triangles) and the pulsed-current (point-up triangles) regime. The dashed line is the calculated variation of the FMR frequency due to the Oersted field of the current. (d) The effective damping constant vs current, determined in the continuous- (point-down triangles) and the pulsed-current (point-up triangles) regimes. The line is the linear fit to the data.

Fig. 2 (a) Schematic of the spin-wave spectrum for an in-plane magnetized ferromagnetic film (right), and a typical BLS spectrum of magnetic fluctuations dominated by the contribution of long-wavelength spin-wave modes (left). The lightly shaded area is the spin-wave manifold, its part contributing to the BLS spectrum is shaded dark. (b) Representative BLS spectra of magnetic fluctuations in the SHE system recorded at different currents in the Pt microstrip, as labeled. (c) Normalized integral intensity under the BLS peak (point-up triangles) and its inverse (point-down triangles) vs current. (d) The FMR frequency $f_0$ (point-up triangles) and the effective magnetization normalized by its value at $I=0$ (point-down triangles) vs current. (e) and (f) Temporal evolution of the normalized effective magnetization (point-down triangles) and the integral BLS intensity (point-up triangles) at $I=25$ and 30 mA, respectively.

Fig. 3 (a) Schematic of the nano-gap spin-Hall nano-oscillators. The inset illustrates the local injection of the electric current and the generation of the spin current inside the gap between the electrodes. (b) BLS spectra in the fluctuation enhancement regime. The arrow marks the new spectral peak that appears at the critical current. (c) BLS spectra of magnetization auto-oscillations. (d) Normalized color-coded spatial map of the auto-oscillation bullet mode. Dashed lines show the contours of the electrodes. (e) The FMR frequency in the Py film (point-down triangles), the auto-oscillation frequency at the onset current (point-up triangles), and the oscillation onset current normalized by its minimum value (diamonds) vs static field.

Fig. 4 (a) Schematic of the nanoconstriction spin-Hall oscillator. Inset shows the calculated electric current distribution in the Pt layer. (b) Representative auto-oscillation spectrum measured by electronic microwave spectroscopy at $I$=3.75 mA. (c) Integral microwave power (point-down triangles and right scale) and spectral linewidth of the auto-oscillation signal (point-up triangles and left scale) vs current. (d) Spectra of magnetization oscillations recorded by BLS at the labeled values of current. (e) Spatial map of the dynamic magnetization recorded by BLS in the auto-oscillation regime (left), and the calculated spatial distribution of the demagnetizing field in the plane of the Py layer (right).

Fig. 5 (a) BLS spectra of spin-current induced magnetic auto-oscillations in the presence of an additional microwave signal, whose frequency $f_{MW}$ is varied from 12 to 16 GHz. (b) Electronically measured spectrum of the free-running auto-oscillations in the absence of an external microwave signal (point-down triangles), and the spectrum of auto-oscillations synchronized with an external microwave signal (point-up triangles). The frequency $\delta f$ is measured relative to the peak center, which coincides with $f_{MW}/2$ in the synchronized regime. (c) Synchronization interval vs the amplitude of the dynamic magnetic field of the microwave current measured in the units of the square root of driving microwave power, at the labeled values of dc current.

Fig. 6 Measurement of the magnons' damping in 20 nm thick YIG films grown by pulsed laser deposition. (a) Derivative of the in-plane FMR absorption peak measured at 6 GHz. (b) Frequency dependence of FMR absorption linewidth: the continuous line is a linear fit corresponding to a Gilbert damping coefficient of $2\times10^{-4}$ [132].

Fig. 7 (a) Study of the spin-pumping in YIG/Pt performed on a series of YIG films with varying thicknesses grown by liquid phase epitaxy. The damping parameter measured in the bare YIG films is plotted in blue. In red is the result when the YIG is covered by 7 nm of Pt. (b) Measurement at 77 K of the inverse spin Hall voltage produced by the FMR excitation of a mm-long slab of 200 nm thick YIG covered by Pt [144]. When a charge current is injected in the adjacent Pt layer, no significant variation of the linewidth is observed for current density up to about $10^{10}$ A/m$^2$. (c) Prediction of the dependence on the YIG thickness of the critical current, $J_C^*$ (onset of damping compensation) according to Eqs. (5) to (7) using the parameters of Table 1.

Fig. 8 (a) Optical image of 20 nm thick YIG disks, with (red) and without (blue) Pt on top, placed below a microwave antenna. Evolution of the FMR linewith when the 20 nm thick YIG film, grown by PLD, is patterned into micro-sized disks : (b) FMR absorption line of the Kittel mode in the extended YIG reference film measured at 8.20GHz; (c) lowest energy mode (coherent mode) of the 700 nm YIG disk measured at 4.99 kOe. The measurements are performed in the normal configuration (perpendicularly magnetized). (d) Comparison of the frequency dependence of the FMR linewidth of the reference YIG film and of the YIG nano-disk (700 nm). The reduction of the linewidth observed in the latter case is due to the vanishing inhomogeneous broadening (zero ordinate intercept) [153].

Fig. 9 ISHE-based FMR spectroscopy of YIG/Pt microdisks. (a) Sketch of the experiment with the micrograph of the device with two microdisks. Scale bar is 50 μm. The in-plane bias field $H$ is oriented at an azimuthal angle φ=90° with respect to the direction of the dc current flowing in the Pt electrode. The excitation field $h_{rf}$ is produced by the rf voltage $V_y$ applied to the inductive loop. The dc voltage $V_x$ across Pt is monitored as a function of the magnetic field strength. (b) ISHE-detected FMR spectra of the 4 μm and 2 μm YIG(20nm)/Pt(8nm) disks at 1 GHz and 4 GHz, respectively. (c) Magnetic-field dependence of frequency of the main FMR mode of the microdiscs. The continuous line is a fit of the experimental data based on the Kittel law [76].

Fig. 10 Evolution of the FMR spectra of a YIG/Pt microdisk (diameter 5μm) as a function of the dc current flowing in Pt measured for $H_0 \parallel +y$ (a) and $H_0 \parallel -y$ (b). The highest-amplitude mode (shaded area) is used for linewidth analysis. Horizontal axes are shifted to align the peaks vertically [61].

Fig. 11 Inductive rf voltage $V_y$ produced in the antenna by auto-oscillations in the 4 μm (a) and 2 μm (b) YIG/Pt disks as a function of the dc current in Pt. The measurements were performed at the bias field $H$ =0.65 kOe. (c)-(e) Current dependences of the auto-oscillation frequency, linewidth, and integrated power, respectively. (f) Frequency dependence of the FMR linewidth in the two microdisks. The continuous lines are linear fits to the experimental data. The dashed line shows the homogeneous contribution in the bare YIG film. (g) Magnetic-field dependence of the threshold current. Expectations taking into account only the homogeneous linewidth or the total linewidth (homogeneous + inhomogeneous) are respectively shown by dashed and continuous lines [76].

Fig. 12 (a) Schematic of the micro-focus BLS experiment with YIG-based SHE system. (b) Typical spatial map of the intensity of current-induced magnetic auto-oscillations in the YIG disk. The contours of the disk and the edges of the electrodes are shown by white lines. The map was recorded at $I$ = 9 mA and $H_0$ = 1000 Oe. (c) Spatial profiles of the oscillation intensity in the direction across and along the gap between the electrodes recorded for different dc currents, as labeled. (d) Current dependence of the full width at half maximum of the spatial profiles along the gap. (e) Current dependences of the maximum intensity of the auto-oscillations detected in the center of the gap (point-down triangles) and that of the spatially-integrated intensity (point-up triangles). (f) Spatial maps of the field-excited ferromagnetic-resonance mode measured at $I$ = 0 and 7 mA, as labeled.

Fig. 13 (a) Schematic of the nonlocal spin-injection nano-oscillator. (b) BLS spectra of magnetization oscillations in the test device recorded at the labeled values of dc current. $H_0$= 750 Oe. (c) The onset current (diamonds and right vertical scale), the uniform FMR frequency (point-down triangles and left vertical scale), and the the auto-oscillation frequency at the oscillation onset (point-up triangles) vs static field. (d) Spatial map of the intensity of magnetization auto-oscillations recorded by BLS at $I$=8 mA and $H_0$=750 Oe. (e) Experimentally determined (symbols) and simulated (curves) frequencies of the fundamental and the secondary auto-oscillation modes vs current. $H_0$= 2000 Oe. (f) Representative spectrum of the electronic signal generated by the nano-oscillator in the inductive antenna. $H_0$= 600 Oe. (g) Calculated maximum increase of temperature in the active device area of the nanoconstriction SHE oscillator (point-down triangles) and the NLSI oscillator (point-up triangles) vs current normalized by the critical current.

Fig. 14 (a) Spatial BLS map of the intensity of overlapping auto-oscillations of two NLSI oscillators separated by the distance of 400 nm. The map was recorded at $I$=20 mA and $H_0$= 1000 Oe. The white circles mark the positions of the nanocontacts. (b) Auto-oscillation frequencies for devices A (point-down triangles) and B (point-up triangles) and their difference (diamonds) vs current. The vertical solid lines mark the values of the current corresponding to the onset ($I_{SYN}$) and lost ($I_L$) of synchronization. (c) Static-field dependencies of the auto-oscillation onset current $I_C$, the synchronization onset current $I_{SYN}$, and the current $I_L$ at which the synchronization is lost.

Fig. 15 (a) Schematic of a hybrid nanomagnonic system enabling excitation of propagating spin waves by pure spin current. The inset illustrates the dipolar field $H_D$ in the stripe waveguide caused by uncompensated magnetic charges at its edges. (b) Solid lines – dispersion spectra of spin waves in an extended Py film with the thickness of 5 and 25 nm, as labeled. Symbols – dispersion spectrum of a spin-wave mode in a 500 nm wide and 20 nm thick stripe waveguide manufactured on top of 5 nm thick Py film. Calculations were performed for $H_0$=1000 Oe. Dashed horizontal line marks the FMR frequency. (c) Current dependences of the auto-oscillation frequency (point-down triangles) and the intensity (point-up triangles) of the NLSI nano-oscillator. $H_0$=1000 Oe. (d) Normalized spatial map of the dynamic magnetization recorded by BLS at $I$=4 mA. Dashed line indicates the contour of the waveguide. (e) Propagation-coordinate dependence of the spin-wave intensity. Solid curve shows the result of the fit of the experimental data (symbols) by the exponential function. Inset – transverse profile of the spin-wave intensity at $x$=2 μm. (f) Dependence of the propagation length of current-induced spin waves and of the coupling coefficient between the NLSI nano-oscillator and the waveguide on the driving current.

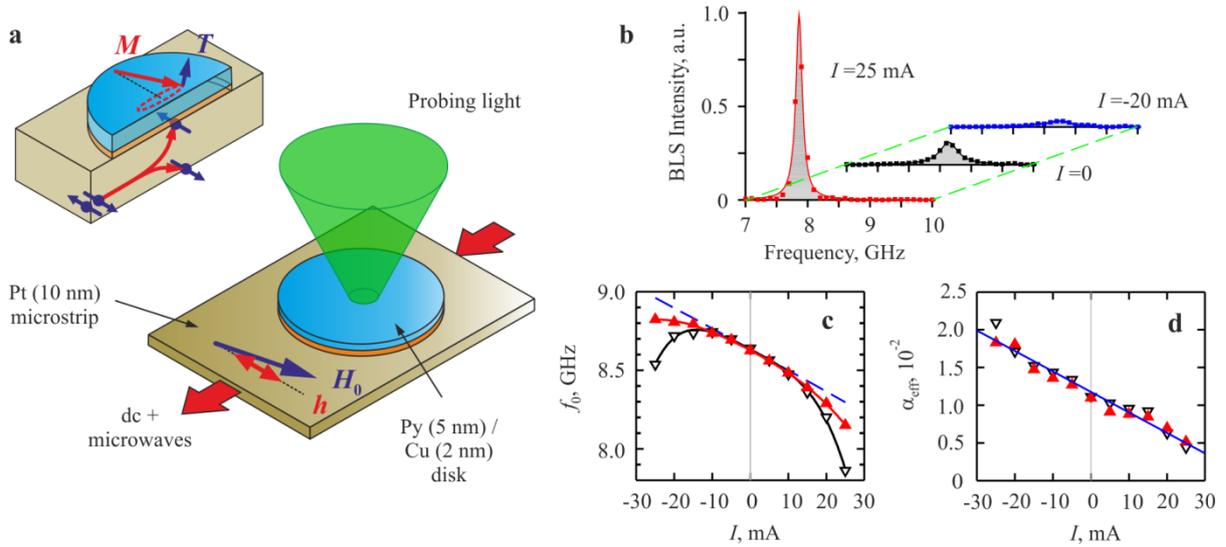

Fig. 1 Schematic of a simple SHE device. Inset illustrates the generation of pure spin current due to SHE in the Pt microstrip. (b) Representative FMR curves recorded by BLS at different currents in the Pt microstrip, as labeled. (c) FMR frequency vs current, measured in the continuous- (point-down triangles) and the pulsed-current (point-up triangles) regime. The dashed line is the calculated variation of the FMR frequency due to the Oersted field of the current. (d) The effective damping constant vs current, determined in the continuous- (point-down triangles) and the pulsed-current (point-up triangles) regimes. The line is the linear fit to the data.

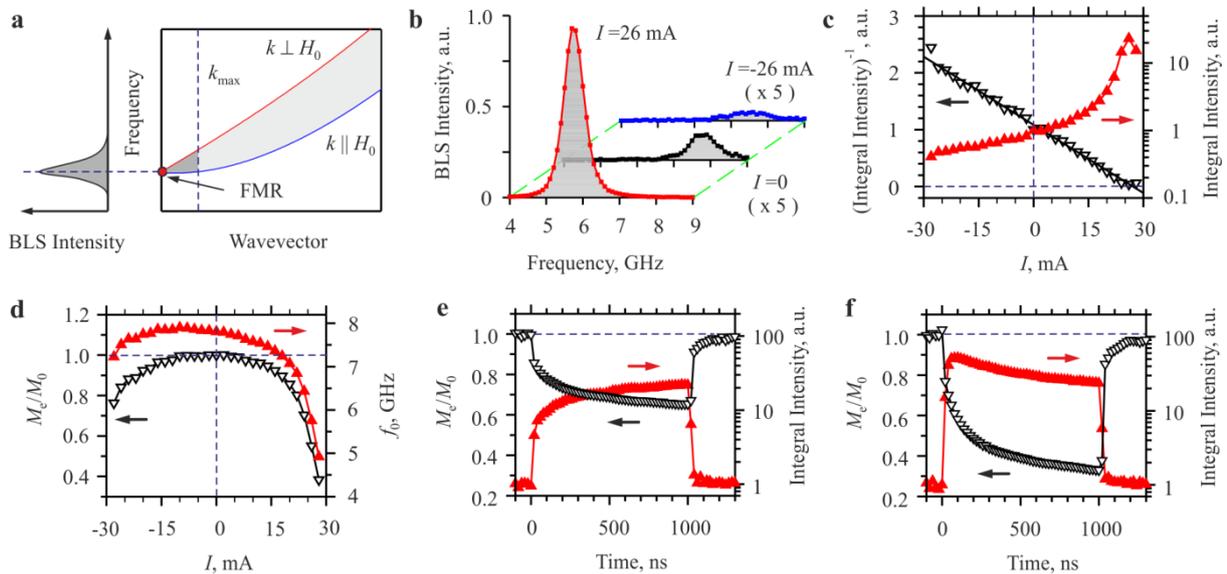

Fig. 2 (a) Schematic of the spin-wave spectrum for an in-plane magnetized ferromagnetic film (right), and a typical BLS spectrum of magnetic fluctuations dominated by the contribution of long-wavelength spin-wave modes (left). The lightly shaded area is the spin-wave manifold, its part contributing to the BLS spectrum is shaded dark. (b) Representative BLS spectra of magnetic fluctuations in the SHE system recorded at different currents in the Pt microstrip, as labeled. (c) Normalized integral intensity under the BLS peak (point-up triangles) and its inverse (point-down triangles) vs current. (d) The FMR frequency $f_0$ (point-up triangles) and the effective magnetization normalized by its value at $I=0$ (point-down triangles) vs current. (e) and (f) Temporal evolution of the normalized effective magnetization (point-down triangles) and the integral BLS intensity (point-up triangles) at $I=25$ and $30$ mA, respectively.

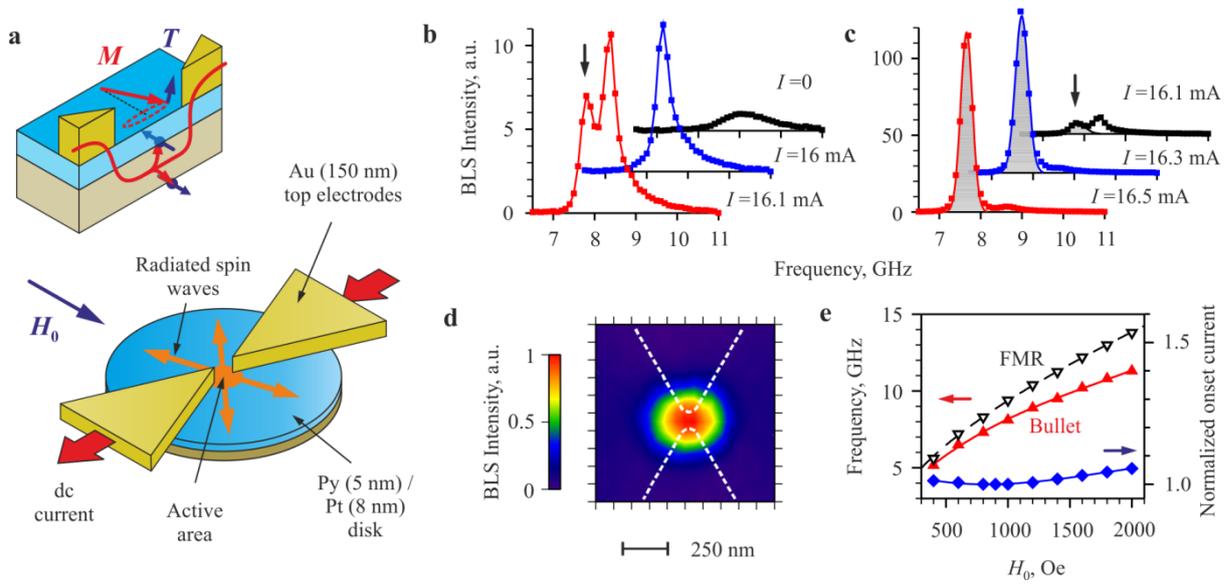

Fig. 3 (a) Schematic of the nano-gap spin-Hall nano-oscillators. The inset illustrates the local injection of the electric current and the generation of the spin current inside the gap between the electrodes. (b) BLS spectra in the fluctuation enhancement regime. The arrow marks the new spectral peak that appears at the critical current. (c) BLS spectra of magnetization auto-oscillations. (d) Normalized color-coded spatial map of the auto-oscillation bullet mode. Dashed lines show the contours of the electrodes. (e) The FMR frequency in the Py film (point-down triangles), the auto-oscillation frequency at the onset current (point-up triangles), and the oscillation onset current normalized by its minimum value (diamonds) vs static field.

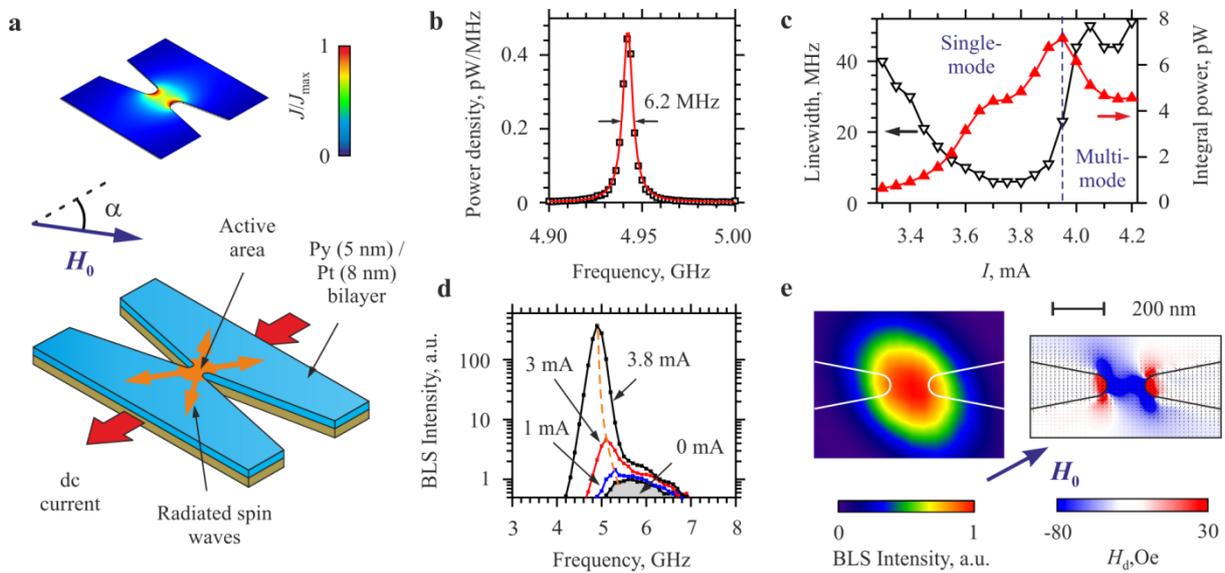

Fig. 4 (a) Schematic of the nanoconstriction spin-Hall oscillator. Inset shows the calculated electric current distribution in the Pt layer. (b) Representative auto-oscillation spectrum measured by electronic microwave spectroscopy at $I$=3.75 mA. (c) Integral microwave power (point-down triangles and right scale) and spectral linewidth of the auto-oscillation signal (point-up triangles and left scale) vs current. (d) Spectra of magnetization oscillations recorded by BLS at the labeled values of current. (e) Spatial map of the dynamic magnetization recorded by BLS in the auto-oscillation regime (left), and the calculated spatial distribution of the demagnetizing field in the plane of the Py layer (right).

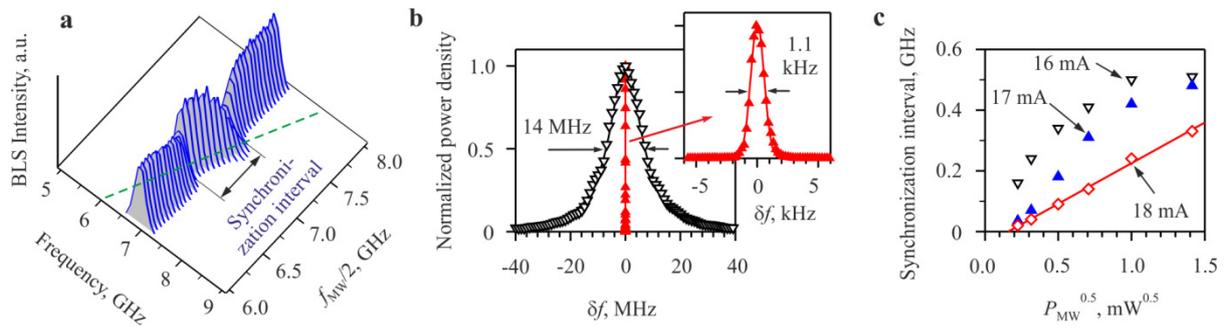

Fig. 5 (a) BLS spectra of spin-current induced magnetic auto-oscillations in the presence of an additional microwave signal, whose frequency $f_{MW}$ is varied from 12 to 16 GHz. (b) Electronically measured spectrum of the free-running auto-oscillations in the absence of an external microwave signal (point-down triangles), and the spectrum of auto-oscillations synchronized with an external microwave signal (point-up triangles). The frequency $\delta f$ is measured relative to the peak center, which coincides with $f_{MW}/2$ in the synchronized regime. (c) Synchronization interval vs the amplitude of the dynamic magnetic field of the microwave current measured in the units of the square root of driving microwave power, at the labeled values of dc current.

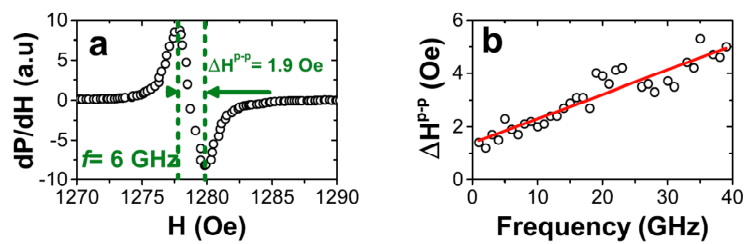

Fig. 6 Measurement of the magnons' damping in 20 nm thick YIG films grown by pulsed laser deposition. (a) Derivative of the in-plane FMR absorption peak measured at 6 GHz. (b) Frequency dependence of FMR absorption linewidth: the continuous line is a linear fit corresponding to a Gilbert damping coefficient of $2\times10^{-4}$ [132].

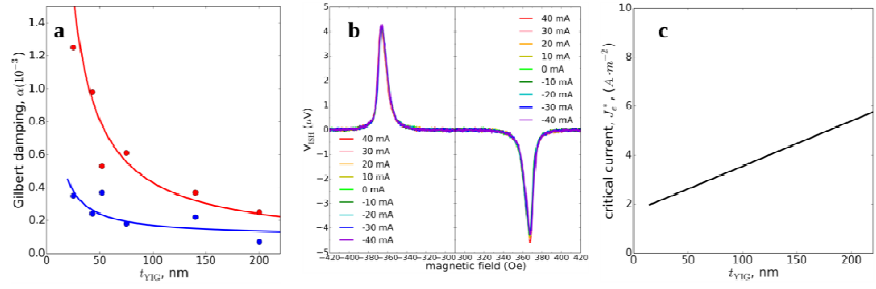

Fig. 7 (a) Study of the spin-pumping in YIG/Pt performed on a series of YIG films with varying thicknesses grown by liquid phase epitaxy. The damping parameter measured in the bare YIG films is plotted in blue. In red is the result when the YIG is covered by 7 nm of Pt. (b) Measurement at 77 K of the inverse spin Hall voltage produced by the FMR excitation of a mm-long slab of 200 nm thick YIG covered by Pt [144]. When a charge current is injected in the adjacent Pt layer, no significant variation of the linewidth is observed for current density up to about $10^{10}$ A/m$^2$. (c) Prediction of the dependence on the YIG thickness of the critical current, $J_C^*$ (onset of damping compensation) according to Eqs. (5) to (7) using the parameters of Table 1.

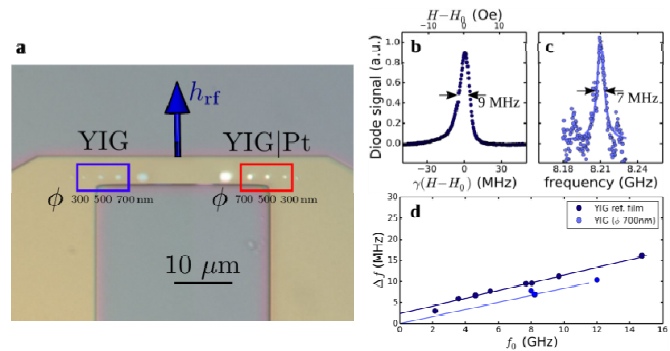

Fig. 8 (a) Optical image of 20 nm thick YIG disks, with (red) and without (blue) Pt on top, placed below a microwave antenna. Evolution of the FMR linewith when the 20 nm thick YIG film, grown by PLD, is patterned into micro-sized disks : (b) FMR absorption line of the Kittel mode in the extended YIG reference film measured at 8.20GHz; (c) lowest energy mode (coherent mode) of the 700 nm YIG disk measured at 4.99 kOe. The measurements are performed in the normal configuration (perpendicularly magnetized). (d) Comparison of the frequency dependence of the FMR linewidth of the reference YIG film and of the YIG nano-disk (700 nm). The reduction of the linewidth observed in the latter case is due to the vanishing inhomogeneous broadening (zero ordinate intercept) [153].

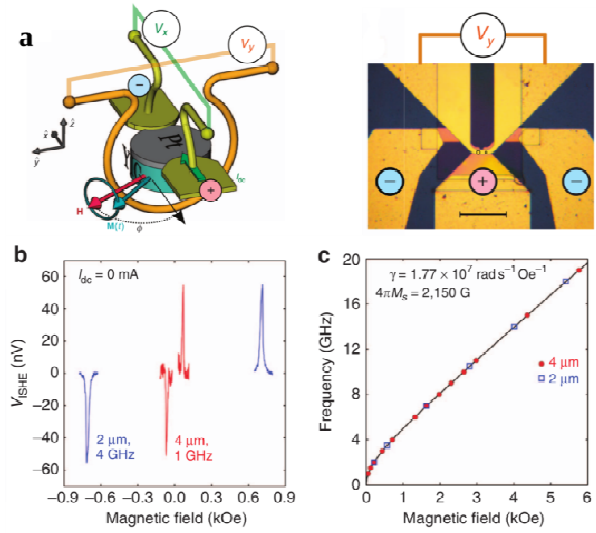

Fig. 9 ISHE-based FMR spectroscopy of YIG/Pt microdisks. (a) Sketch of the experiment with the micrograph of the device with two microdisks. Scale bar is 50 µm. The in-plane bias field $H$ is oriented at an azimuthal angle φ=90° with respect to the direction of the dc current flowing in the Pt electrode. The excitation field $h_{rf}$ is produced by the rf voltage $V_y$ applied to the inductive loop. The dc voltage $V_x$ across Pt is monitored as a function of the magnetic field strength. (b) ISHE-detected FMR spectra of the 4 µm and 2 µm YIG(20nm)/Pt(8nm) disks at 1 GHz and 4 GHz, respectively. (c) Magnetic-field dependence of frequency of the main FMR mode of the microdiscs. The continuous line is a fit of the experimental data based on the Kittel law [76].

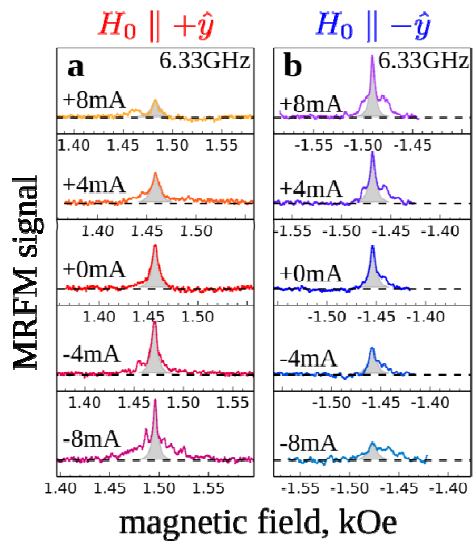

Fig. 10 Evolution of the FMR spectra of a YIG/Pt microdisk (diameter 5µm) as a function of the dc current flowing in Pt measured for $H_0 \parallel +y$ (a) and $H_0 \parallel -y$ (b). The highest-amplitude mode (shaded area) is used for linewidth analysis. Horizontal axes are shifted to align the peaks vertically [61].

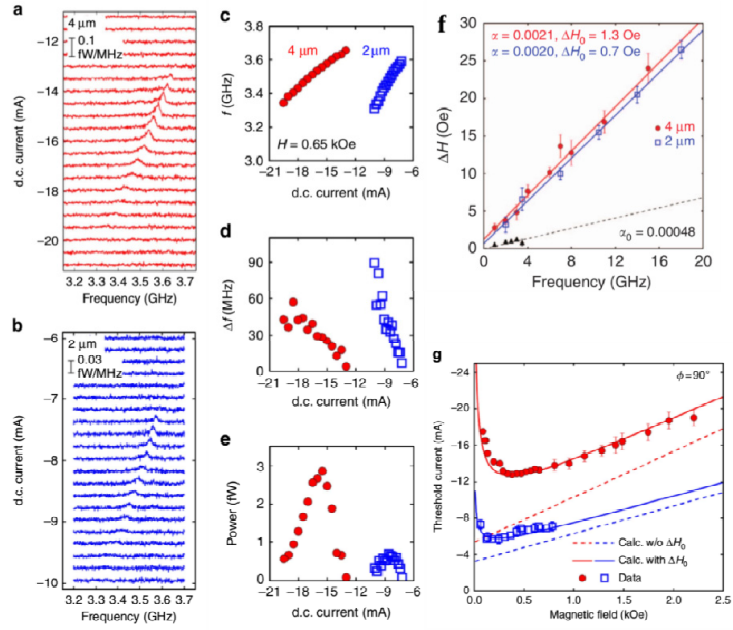

Fig. 11 Inductive rf voltage $V_y$ produced in the antenna by auto-oscillations in the 4 µm (a) and 2 µm (b) YIG/Pt disks as a function of the dc current in Pt. The measurements were performed at the bias field $H$ =0.65 kOe. (c)-(e) Current dependences of the auto-oscillation frequency, linewidth, and integrated power, respectively. (f) Frequency dependence of the FMR linewidth in the two microdisks. The continuous lines are linear fits to the experimental data. The dashed line shows the homogeneous contribution in the bare YIG film. (g) Magnetic-field dependence of the threshold current. Expectations taking into account only the homogeneous linewidth or the total linewidth (homogeneous + inhomogeneous) are respectively shown by dashed and continuous lines [76].

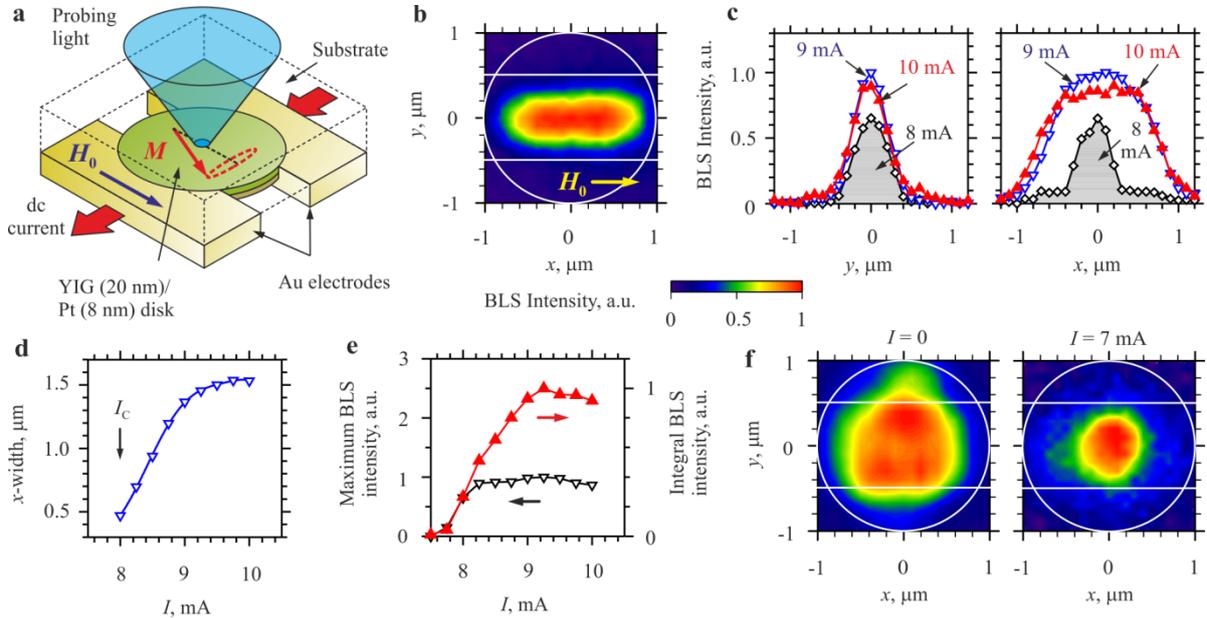

Fig. 12 (a) Schematic of the micro-focus BLS experiment with YIG-based SHE system. (b) Typical spatial map of the intensity of current-induced magnetic auto-oscillations in the YIG disk. The contours of the disk and the edges of the electrodes are shown by white lines. The map was recorded at $I$ = 9 mA and $H_0$ = 1000 Oe. (c) Spatial profiles of the oscillation intensity in the direction across and along the gap between the electrodes recorded for different dc currents, as labeled. (d) Current dependence of the full width at half maximum of the spatial profiles along the gap. (e) Current dependences of the maximum intensity of the auto-oscillations detected in the center of the gap (point-down triangles) and that of the spatially-integrated intensity (point-up triangles). (f) Spatial maps of the field-excited ferromagnetic-resonance mode measured at $I$ = 0 and 7 mA, as labeled.

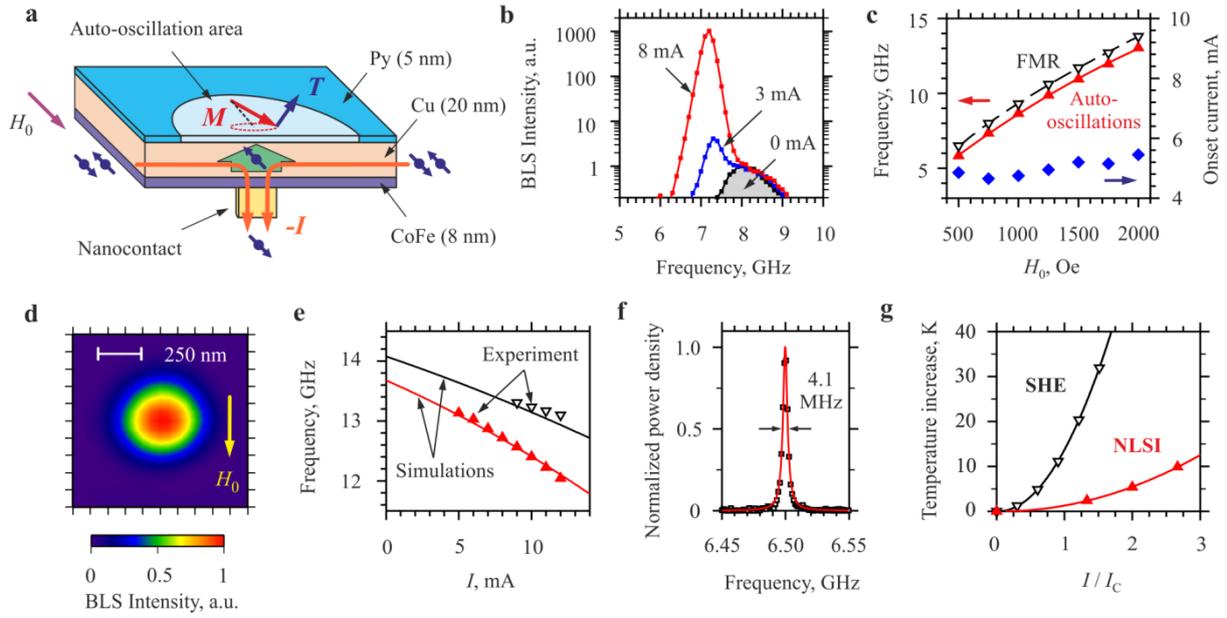

Fig. 13 (a) Schematic of the nonlocal spin-injection nano-oscillator. (b) BLS spectra of magnetization oscillations in the test device recorded at the labeled values of dc current. $H_0$= 750 Oe. (c) The onset current (diamonds and right vertical scale), the uniform FMR frequency (point-down triangles and left vertical scale), and the the auto-oscillation frequency at the oscillation onset (point-up triangles) vs static field. (d) Spatial map of the intensity of magnetization auto-oscillations recorded by BLS at $I$=8 mA and $H_0$= 750 Oe. (e) Experimentally determined (symbols) and simulated (curves) frequencies of the fundamental and the secondary auto-oscillation modes *vs* current. $H_0$= 2000 Oe. (f) Representative spectrum of the electronic signal generated by the nano-oscillator in the inductive antenna. $H_0$= 600 Oe. (g) Calculated maximum increase of temperature in the active device area of the nanoconstriction SHE oscillator (point-down triangles) and the NLSI oscillator (point-up triangles) vs current normalized by the critical current.

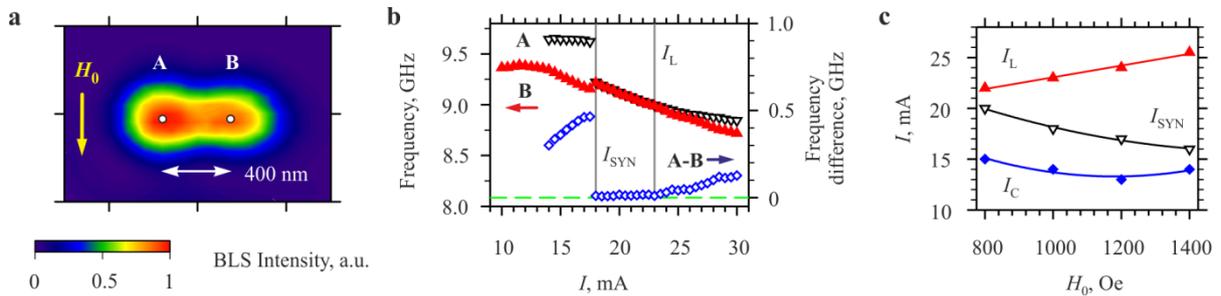

Fig. 14 (a) Spatial BLS map of the intensity of overlapping auto-oscillations of two NLSI oscillators separated by the distance of 400 nm. The map was recorded at $I$=20 mA and $H_0$= 1000 Oe. The white circles mark the positions of the nanocontacts. (b) Auto-oscillation frequencies for devices A (point-down triangles) and B (point-up triangles) and their difference (diamonds) vs current. The vertical solid lines mark the values of the current corresponding to the onset ($I_{SYN}$) and lost ($I_L$) of synchronization. (c) Static-field dependencies of the auto-oscillation onset current $I_C$, the synchronization onset current $I_{SYN}$, and the current $I_L$ at which the synchronization is lost.

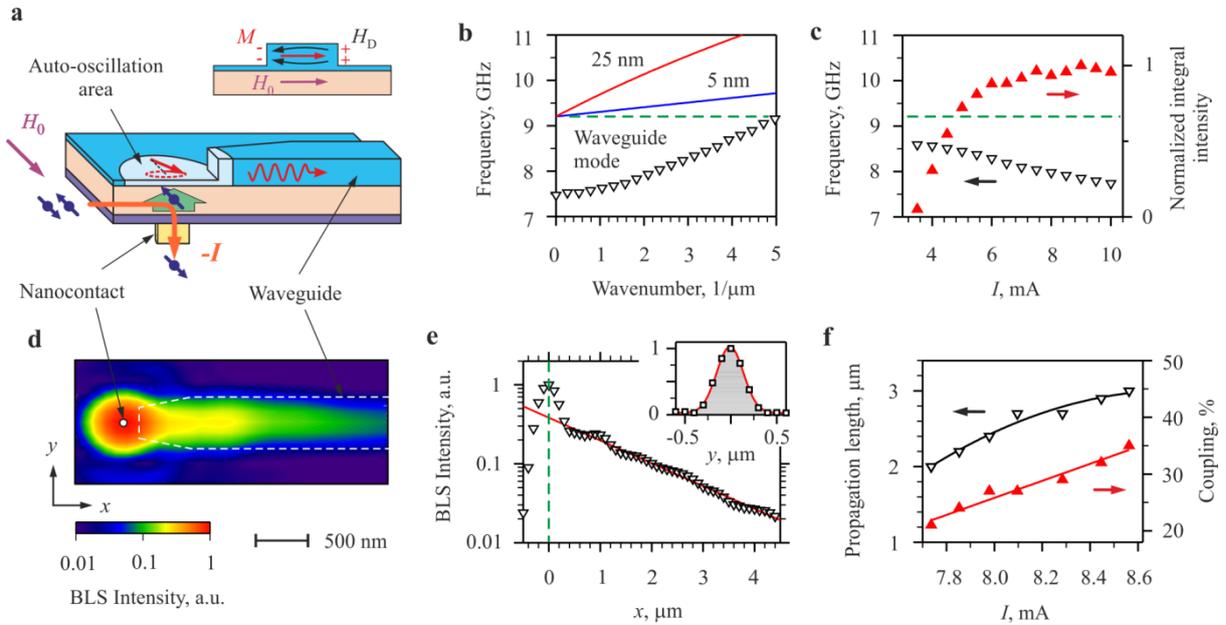

Fig. 15 (a) Schematic of a hybrid nanomagnonic system enabling excitation of propagating spin waves by pure spin current. The inset illustrates the dipolar field $H_D$ in the stripe waveguide caused by uncompensated magnetic charges at its edges. (b) Solid lines – dispersion spectra of spin waves in an extended Py film with the thickness of 5 and 25 nm, as labeled. Symbols – dispersion spectrum of a spin-wave mode in a 500 nm wide and 20 nm thick stripe waveguide manufactured on top of 5 nm thick Py film. Calculations were performed for $H_0$=1000 Oe. Dashed horizontal line marks the FMR frequency. (c) Current dependences of the auto-oscillation frequency (point-down triangles) and the intensity (point-up triangles) of the NLSI nano-oscillator. $H_0$=1000 Oe. (d) Normalized spatial map of the dynamic magnetization recorded by BLS at $I$=4 mA. Dashed line indicates the contour of the waveguide. (e) Propagation-coordinate dependence of the spin-wave intensity. Solid curve shows the result of the fit of the experimental data (symbols) by the exponential function. Inset – transverse profile of the spin-wave intensity at $x$=2 µm. (f) Dependence of the propagation length of current-induced spin waves and of the coupling coefficient between the NLSI nano-oscillator and the waveguide on the driving current.